\documentclass[twocolumn,
                showpacs,
                nofootinbib,
                nobibnotes,
                aps,
                superscriptaddress,
                amssymb,
                amsmath,
                floatfix,
                reprint,
                prx,
                notitlepage,
                showkeys,
                10pt,
                longbibliography,
		]{revtex4-1}

\usepackage{amssymb}
\usepackage{amsmath,amsbsy}
\usepackage{mathtools}  
\DeclareMathOperator{\Tr}{Tr} 
\usepackage{cprotect}
\usepackage{graphicx}
\usepackage{dcolumn}
\usepackage{url}
\usepackage[colorlinks=true,breaklinks=true,allcolors=blue]{hyperref}
\usepackage{tikz}
\usepackage{dsfont}
\usepackage{epsfig}
\usepackage{color}
\usepackage[ruled]{algorithm2e}
\usepackage{bbold}
\usepackage{glossaries}
\usepackage{xcolor}
\usepackage{tikz}
\usepackage{bm}
\usepackage{hhline}
\usepackage{spverbatim}
\usepackage{siunitx}
\usepackage{braket} 
\usepackage{chngcntr}
\makeatletter
\let\cat@comma@active\@empty
\makeatother

\begin{document}
\title{The statistical mechanics and machine learning of the $\alpha$-Rényi ensemble}

\author{Andrew Jreissaty}
\email{ajreissaty@phys.ethz.ch}
\affiliation{Institute for Theoretical Physics, ETH Zurich, CH-8093 Zurich, Switzerland}
\affiliation{Vector Institute, MaRS Centre, Toronto, Ontario, M5G 1M1, Canada}
\affiliation{Department of Physics and Astronomy, University of Waterloo, Ontario, N2L 3G1, Canada}

\author{Juan Carrasquilla}
\affiliation{Institute for Theoretical Physics, ETH Zurich, CH-8093 Zurich, Switzerland}
\affiliation{Vector Institute, MaRS Centre, Toronto, Ontario, M5G 1M1, Canada}
\affiliation{Department of Physics and Astronomy, University of Waterloo, Ontario, N2L 3G1, Canada}

\date{April 18, 2024}

\begin{abstract}
We study the statistical physics of the classical Ising model in the so-called $\alpha$-Rényi ensemble, a finite-temperature thermal state approximation that minimizes a modified free energy based on the $\alpha$-Rényi entropy. We begin by characterizing its critical behavior in mean-field theory in different regimes of the Rényi index $\alpha$. Next, we re-introduce correlations and consider the model in one and two dimensions, presenting analytical arguments for the former and devising a Monte Carlo approach to the study of the latter. Remarkably, we find that while mean-field predicts a continuous phase transition below a threshold index value of $\alpha \sim 1.303$ and a first-order transition above it, the Monte Carlo results in two dimensions point to a continuous transition at all $\alpha$. We conclude by performing a variational minimization of the $\alpha$-Rényi free energy using a recurrent neural network (RNN) ansatz where we find that the RNN performs well in two dimensions when compared to the Monte Carlo simulations. Our work highlights the potential opportunities and limitations associated with the use of the $\alpha$-Rényi ensemble formalism in probing the thermodynamic equilibrium properties of classical and quantum systems.
\end{abstract}

\maketitle

\section{Introduction} \label{sec:introduction_final_paper}

Simulating finite-temperature states, both in equilibrium and out-of-equilibrium, remains a significant challenge in the study of quantum many-body systems. Quantum Monte Carlo approaches, long considered state-of-the art for the simulation of equilibrium states in quantum many-body systems, are plagued by fundamental sign problem issues in fermionic and frustrated quantum spin systems \cite{ScalettarQMC,SandvikQMC_1,SandvikQMC_2,TroyerQMC,WuQMC,LiQMC,WeiQMC,Rubem2022,Broecker2016,Mak1990}. More recently, a large number of approaches for Gibbs state simulation involving the imaginary time evolution of a purified mixed state to produce thermal pure quantum states (TPQS) have been introduced~\cite{TPQS1_2012,TPQS2_2013,TPQS3_2016,TPQS4_2021,Nomura2021,NNQS_FiniteT_IrikuraSaito2020}. Other approaches such as minimally entangled typical thermal states have also been proposed, leveraging matrix product state (MPS) algorithms along the way \cite{StevenWhite2009,StoudenmireWhite2010}. In time, many in the community have turned to the variational method, proposing TPQS and density matrix ansätze parameterized by a set of parameters that are tuned to approximate the Liouvillian dynamics of mixed states coupled to Markovian baths using the time-dependent variational principle\cite{JannesZakari2023,FilippoGramHadamard2022}. Those behind the vast majority of these approaches have recognized a common issue: simulating the Gibbs state variationally by minimizing the Gibbs free energy at finite temperature is challenging due to the issues associated with computing the von Neumann entropy of a parameterized quantum density matrix. As such, thermal state approximations have started to emerge. One such approximation involves the minimization of a modified free energy known as the 2-Rényi free energy, where the von Neumann entropy is replaced by the second Rényi entropy \cite{Bashkirov2004}. In this way, the 2-Rényi ensemble, which minimizes the 2-Rényi free energy, has provided a fresh breeding ground for quantum simulation of finite-temperature states, in particular using MPS and neural network quantum state (NNQS) approaches. 

In the last few years, machine learning and NNQS models ranging from restricted Boltzmann machines (RBMs) to convolutional neural networks (CNNs) among others have exploded onto the scene, providing highly expressive variational ansätze as well as other techniques for the efficient simulation of ground state wavefunctions, the detection of continuous phase transitions and the reconstruction of quantum states
\cite{CarleoTroyer2017,JuanRoger2017,Magnifico2023,Dong2019,Zhang2019,Uvarov2020,Che2020,Wang2016,Canabarro2019,Hu2017,CarleoCiracCoReview2019,MelkoCarleoCarrasquillaCiracRBM2019,TorlaiTomography2018,Moritz2022, Raghu2017}. The continued development of NNQS has since resulted in the emergence of a highly efficient autoregressive model based on recurrent neural networks (RNNs) which has been used for ground state wavefunction optimization in both frustrated and unfrustrated spin systems as well as Fermionic systems \cite{MohamedRNN2020,MohamedRoelandAnnealing2021,MohamedTopologicalOrder2023,langeNeuralNetworkApproach2024}. Some studies have sought to enhance RNN ground state optimizations by leveraging quantum simulation and Monte Carlo sampling data in the process \cite{StefanieSchuyler2022,Schuyler2024}, demonstrating the flexibility of the overall NNQS approach.

Although the work in Refs.~\cite{Bashkirov2004,Giudice2021,Giudice2024} has focused on studying finite-temperature properties of quantum systems through the Rényi ensemble, here we take a step back and examine whether the Rényi ensemble provides an accurate approximation of the Gibbs state at the classical level.  We focus on the Ising model in one and two dimensions~\cite{Onsager1944}, which, in light of its analytical and numerical tractability, provides an ideal playground for understanding to what extent and in which regimes the $\alpha$-Rényi ensemble reproduces the physics of the Gibbs state.  We first consider a mean-field solution of the model within the ensemble, followed by a detailed exploration of the model in the presence of fluctuations through the development of a Markov-chain Monte Carlo (MCMC) technique specifically designed to target the Rényi ensemble. In the latter case, sampling via Monte Carlo presents a challenge as the distribution itself depends on the average energy, which we estimate via an iterative procedure. Beyond our Monte Carlo approach, we consider variational approximations to the Rényi ensemble using recurrent neural networks and assess their quality by comparing their output to Monte Carlo and exact approaches.

The paper is organized as follows. In Sec.~\ref{sec:the_Renyi_ensemble_final_paper}, we introduce the Rényi ensemble, which is the foundation of paper. Next, we vet the Rényi ensemble approximation by applying it to the mean-field study of the Ising model in Sec.~\ref{sec:mean-field_final_paper}, followed by an analytical treatment of the one-dimensional (1D) Ising model in this ensemble in Sec.~\ref{sec:1D_exact_solution_final_paper}. We re-introduce correlations in Sec.~\ref{sec:Monte_Carlo}, presenting Monte Carlo results for the two-dimensional  (2D) Ising model in the Rényi ensemble, and in Sec.~\ref{sec:RNN} we compare those results with the RNN predictions. In Sec.~\ref{sec:Conclusion}, we conclude and discuss the future outlook of our work and thermal state approximations more broadly.

\section{The Rényi Ensemble} \label{sec:the_Renyi_ensemble_final_paper}
We consider the $\alpha$-Rényi free energy \cite{Giudice2021} given by
\begin{equation} \label{eq:alphaRenyiFreeEnergy}
    F_{\alpha} = \Tr(\rho H) - T \, \frac{1}{1-\alpha} \log\left[\Tr\left(\rho^\alpha\right)\right].
\end{equation}
Here, $\alpha\ge 1$ is the Rényi index and $\rho$ is the density matrix of the system. The $\alpha$-Rényi ensemble is defined as the density matrix $\rho^{(\alpha)}$ that minimizes $F_{\alpha}$. It is expressed as $\rho^{(\alpha)} = \sum_k p_k^{(\alpha)} \sum_{j_k=1}^{N_k} \ket{E_k^{(j_k)}}\bra{E_k^{(j_k)}}$, as previously derived in Ref.~\cite{Bashkirov2004} and extensively explored in Refs.~\cite{Giudice2024, Giudice2021}. The eigenstates of the Hamiltonian $\{\ket{E_k^{(j_k)}}\}$ have corresponding $k^\text{th}$ energy levels $E_k$ with degeneracies $N_k$, with the index $j_k$ specifying a particular degenerate state in the $k^\text{th}$ level. The probabilities $p_k^{(\alpha)} \in [0,1]$ are given by
\begin{align} \label{eq:alphaRenyiEnsProb}
    p_k^{(\alpha)} &= \frac{\left[1 - \beta \, \frac{\alpha-1}{\alpha}(E_k - \Bar{E})\right]^\frac{1}{\alpha-1}}{Z_\alpha}, \\ \label{eq:alphaRenyiPartFunc}
    Z_\alpha &= \sum_{k=0}^{n_\beta-1} N_k \left[1 - \beta \, \frac{\alpha-1}{\alpha}(E_k - \Bar{E})\right]^\frac{1}{\alpha-1},\\ \label{eq:alphaRenyiEnsConstr}
    \text{with } & E_k \le \frac{\alpha}{\beta (\alpha-1)} + \Bar{E},
\end{align}
 where the temperature of the ensemble is $T$ and its inverse is $\beta=1/T$.  The probabilities $p_k^{(\alpha)}$ satisfy the constraint $\Tr(\rho^{(\alpha)} H) \equiv \Bar{E}$, where $\Bar{E} = \langle H \rangle$ is the average energy of the system, and $Z_\alpha$ is the partition function of the generalized ensemble. The condition in Eq.~(\ref{eq:alphaRenyiEnsConstr}) must be satisfied in order to ensure $p_k^{(\alpha)}\ge 0$, or in other words the positive semi-definiteness of $\rho^{(\alpha)}$. It is possible to show that in the limit $\alpha \rightarrow 1$, the Rényi ensemble tends exactly to the Gibbs state \cite{AlfredRenyiProceedings}. The sum in Eq.~(\ref{eq:alphaRenyiPartFunc}) is over all $n_\beta$ energy levels that satisfy Eq.~(\ref{eq:alphaRenyiEnsConstr}), a number that depends on the inverse temperature $\beta$.

The average energy $\Bar{E}$ is computed by solving the fixed point equation
\begin{equation} \label{eq:fixed_point_equation}
    \Tr(\rho^{(\alpha)} H) = \Bar{E} \longrightarrow \sum_{k=0}^{n_\beta-1} N_k E_k p_k^{(\alpha)} = \Bar{E}.
\end{equation}
As noted in App.~\ref{appendix:attractive_fixed_point},  we observe that the $\Tr(\rho^{(\alpha)} H) = \Bar{E}$ fixed point is attractive for all $\alpha > 1$ for the 1D and 2D Ising models with no external field. This feature proves especially useful in numerical simulations as it enables the possibility to find $\Bar{E}$ iteratively, which we use for both exact and Monte Carlo simulations of the Ising model.

\section{Ising Model: Mean-Field} \label{sec:mean-field_final_paper}
We first consider the $\alpha$-Rényi ensemble within mean-field theory, focusing on the classical Ising model $H = -J\sum_{<i,j>} \sigma_i \sigma_j$ with $J>0$. Our mean-field calculation follows the approach in Ref.~\cite{Arovas}, which is based on a factorized density matrix 
\begin{equation} \label{eq:rho_Arovas}
    \rho = \bigotimes_{i=1}^N \rho_i \equiv \bigotimes_{i=1}^N \begin{bmatrix}
        \frac{1+m}{2} & 0 \\
    0 & \frac{1-m}{2}
  \end{bmatrix}_i.
\end{equation}
We minimize the resulting $\alpha$-Rényi free energy with respect to the variational parameter $m$. Restricting $m$ to the interval $[-1,1]$ allows for the interpretation of $\rho_i$ as a classical probability distribution over the binary spin values $\{+1,-1\}$ such that the average spin value is $m$. This product state approach is equivalent to other mean-field formulations and can be shown to recover the mean-field equation for the magnetization $m = \tanh{\left[m(qJ)/T\right]}$ of the Ising model in the Gibbs state, with $q = 2d$ the coordination number associated with a hypercubic lattice in $D$ dimensions. Applying Eq.~(\ref{eq:rho_Arovas}) to the $\alpha$-Rényi free energy leads to a free energy per spin of
\begin{align} \label{eq:free_energy_Renyi_Arovas_final_paper}
\begin{split}
    f_\alpha (m) & = -\frac{1}{2}(qJ)m^2 \\
    & - T\,\frac{1}{1-\alpha}\log\left[\left(\frac{1+m}{2}\right)^\alpha + \left(\frac{1-m}{2}\right)^\alpha\right].
\end{split}
\end{align}

\begin{figure}
\includegraphics[scale=0.53]{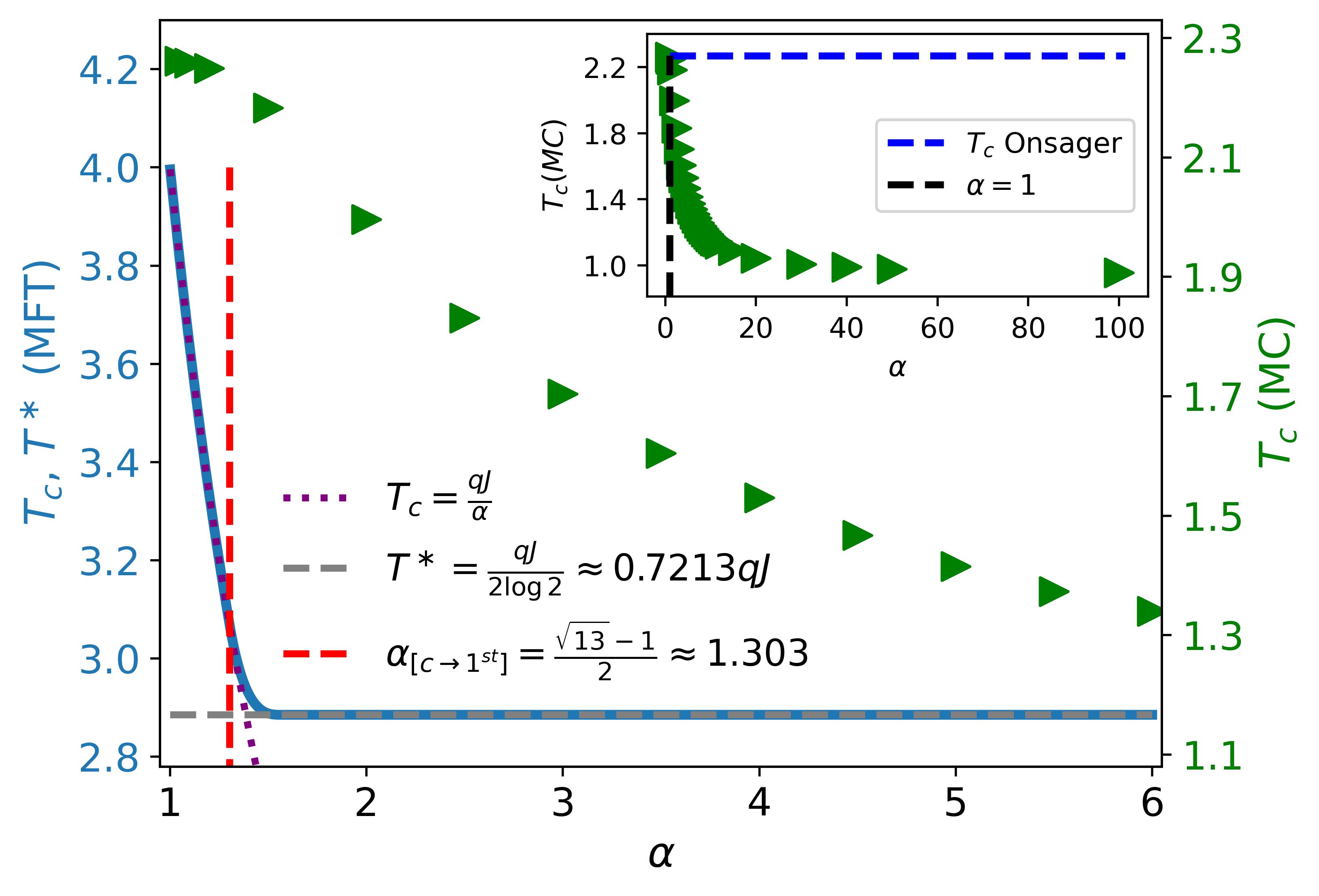}
\caption{\label{fig:epsart} Mean-field critical temperature $T_c$ and transition temperature $T^\ast$ (solid blue), as well as $T_c$ as extracted from Monte Carlo simulations (green), as a function of the Rényi index $\alpha$, for the 2D Ising model ($q=4$). The mean-field data is extracted numerically from Eq.~(\ref{eq:free_energy_Renyi_Arovas_final_paper}), while the Monte Carlo $T_c$ values are computed by data collapse (Sec.~\ref{sec:Monte_Carlo}). The Monte Carlo error bars are smaller than the size of the data points. For mean-field we show the exact functional form for $T_c$ in the continuous regime (purple dotted curve $T_c = qJ / \alpha$), the constant value of $qJ / (2\log 2)$ that $T^\ast$ takes for all $\alpha > \Bar{\alpha} \sim 1.56$ (grey dashed horizontal line) and the threshold $\alpha_{[c\rightarrow 1^\text{st}]}$ (red dashed vertical line) that separates the continuous and first-order regimes. $J$ has been set to 1. The limit $\lim\limits_{\alpha\rightarrow 1} T_c = qJ$ recovers the mean-field result for the Ising model in the Gibbs state. The inset displays Monte Carlo $T_c$ data for large Rényi index, where $T_c$ is seen to approach $T_c \sim 1$ as $\alpha\rightarrow \infty$.}
\label{fig:Tc_vs_alpha}
\end{figure}

\begin{figure}
\includegraphics[scale=0.58]{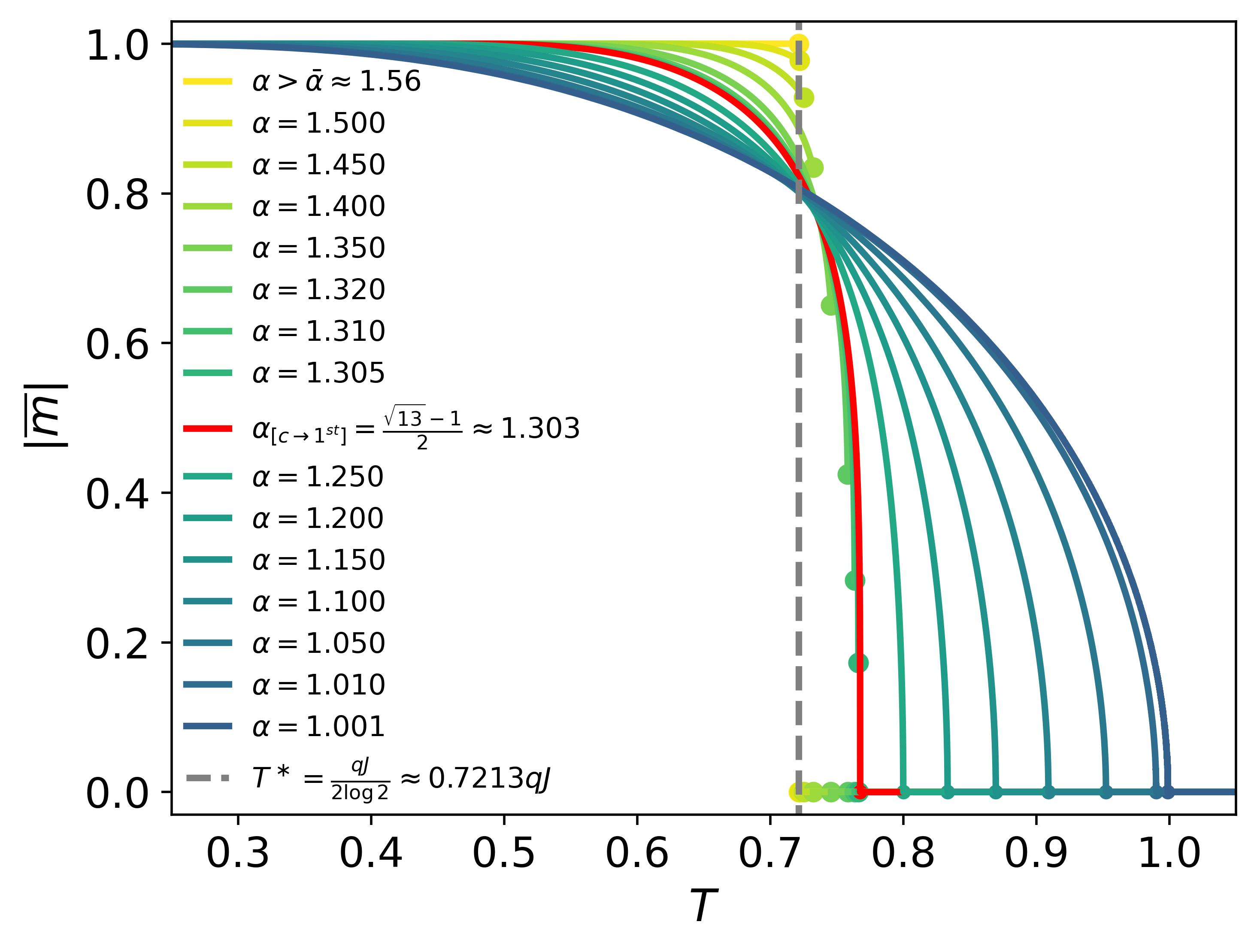}
\caption{\label{fig:epsart} Absolute value of the mean-field magnetization per spin $\overline{m}$ that minimizes $f_\alpha (m)$ as a function of $T$ for various values of $\alpha$. For all $\alpha > \bar{\alpha}\sim 1.56$, the magnetization jump $m_\text{gap}$ is $1$, while for $\alpha \in \left(\frac{\sqrt{13}-1}{2},\sim 1.56\right)$, $m_\text{gap} < 1$. The red curve depicts the magnetization at exactly $\alpha_{[c\rightarrow \text{1st}]}= \frac{\sqrt{13}-1}{2}$. Below this value, the curves are characteristic of a continuous transition. As $\alpha \rightarrow 1$, the critical temperature tends to the Gibbs ensemble prediction $T_c=qJ$. We have set $qJ=1$ for simplicity.}
\label{fig:m_vs_T_final_paper}
\end{figure}

Remarkably, while the mean-field free energy in the Rényi index interval $1\le\alpha\lesssim 1.3$ predicts a continuous phase transition for the Ising model, for $\alpha \gtrsim1.3$ the transition is first-order. This can be seen in Fig.~\ref{fig:mean_field_four_into_one_plot} (see App.~\ref{appendix:mean_field_details}) and Fig.~\ref{fig:m_vs_T_final_paper}, where the hallmarks of continuous and first-order transitions emerge for different values of $\alpha$. This stands in contrast with the well-known case of the Ising model in the Gibbs state ($\alpha\rightarrow 1$), where the mean-field transition is continuous with a critical temperature $T_c = qJ$. The appearance of a first-order transition at larger $\alpha$ arises because as $T$ increases, higher energy states can be "suddenly" turned on and made accessible to the system due to the nature of the Rényi constraint (see RHS of Eq.~(\ref{eq:alphaRenyiEnsConstr})), which hints at the possibility of a discontinuous jump in the value of the mean-field order parameter at some transition temperature $T^\ast$. On the other hand, values of $\alpha$ closer to 1 produce a continuous transition since the Rényi ensemble tends to the Gibbs state as $\alpha\rightarrow 1$. We can see from  Eq.~(\ref{eq:alphaRenyiEnsConstr}) that as $\alpha$ approaches 1, more and more higher energy states are rendered accessible to the system at any given temperature, making discontinuous jumps in $m$ less likely at the mean-field level.

We now derive expressions for the critical temperature $T_c$ in the continuous regime (the $\alpha$ region for which the mean-field transition is continuous) and the transition temperature $T^\ast$ in the first-order regime (the $\alpha$ region for which the mean-field transition is first-order). In particular, we focus on their dependence on the Rényi index $\alpha$. In between, we also derive the value of $\alpha$ that exactly separates the two regimes, which we denote $\alpha_{[c\rightarrow 1^\text{st}]}$.  Let us assume that $\alpha$ is such that the mean field $\alpha$-Rényi free energy in Eq.~(\ref{eq:free_energy_Renyi_Arovas_final_paper}) describes a continuous transition. Then $T_c$ is the temperature at which the nature of the extremum at $m=0$ changes from a local maximum to the global minimum. To derive it, we compute $\frac{\partial^2 f_\alpha}{\partial m^2}\big|_{m=0}$, set it to $0$ and solve for $T_c$. We find
\begin{equation} \label{eq:Tc_vs_alpha_expression}
    \frac{\partial^2 f_\alpha}{\partial m^2}\bigg|_{m=0,T=T_c}  = -qJ + \alpha T_c = 0 \longrightarrow T_c = \frac{qJ}{\alpha}.
\end{equation}
This expression recovers the Gibbs state mean-field limit $\lim_{\alpha \rightarrow 1} T_c = \lim_{\alpha \rightarrow 1} qJ/\alpha = qJ$.

Eq.~(\ref{eq:Tc_vs_alpha_expression}) is valid for $\alpha \in \left[1,\alpha_{[c\rightarrow 1^\text{st}]}\right]$, i.e. the continuous regime of $\alpha$ values. It is possible to evaluate $\alpha_{[c\rightarrow 1^\text{st}]}$ exactly. The procedure involves computing the Taylor expansion of $f_\alpha$ about $m=0$ to $6^\text{th}$ order in $m$, which we denote as $f_\alpha^{(6)}$, extremizing the result, and subsequently identifying the regime of $\alpha$ values for which $f_\alpha^{(6)}$ allows for the possibility of five real extrema depending on the temperature $T$, which is a hallmark of a first-order transition. The reason we conduct this analysis to only $\mathcal{O}(m^6)$ and not greater is that $f_\alpha^{(6)}$ captures the macroscopic "extremal shape" of the true free energy $f_\alpha$ when $f_\alpha$ has five extrema. In other words, whenever $f_\alpha$ has five extrema in the interval $m\in [-1,1]$, $f_\alpha^{(6)}$ also has five extrema, although for the latter, the interval may have to be widened to observe them all depending on the specific Rényi index under consideration. As such, a higher order analysis is not needed. The details of the procedure are laid out in App.~\ref{appendix:mean_field_details}. It finds
\begin{equation} \label{eq:alpha_cont_to_first_order}
    \alpha_{[c\rightarrow 1^\text{st}]} = \frac{\sqrt{13}-1}{2} \approx 1.303,
\end{equation}
which, as opposed to $T_c$, is independent of the dimensionality of the system. In summary, mean-field theory predicts a continuous symmetry-breaking phase transition for the Ising model in the $\alpha$-Rényi ensemble for $\alpha \in \left[1,\frac{\sqrt{13}-1}{2}\right]$ and a first-order transition for $\alpha \in \left(\frac{\sqrt{13}-1}{2},\infty\right)$.

We now focus on the dependence of the first-order transition temperature $T^\ast$ on $\alpha$. By examining the dependence of $f_\alpha$ on $m$ at various values of $\alpha$ in the first-order regime, we find that for $\alpha$ greater than or equal to some value $\bar{\alpha}$, $f_\alpha$ is globally minimized in the interval $m\in[-1,1]$ at $m=\pm 1$ or $m=0$, depending on the temperature, meaning that the jump in magnetization as $T^\ast$ is crossed is exactly $m_\text{gap}=1$ for all $\alpha \ge \bar{\alpha}$. Thus, the transition temperature for all $\alpha \ge \Bar{\alpha}$ can be derived by setting $f_\alpha(m=\pm 1) = f_\alpha(m=0) \nonumber$, which leads to
\begin{equation} \label{eq:T_star_asymptote}
    -\frac{1}{2} (qJ) = -T \frac{1}{1-\alpha} \log \left(2^{1-\alpha}\right) \longrightarrow T^\ast = \frac{qJ}{2\log 2}.
\end{equation}

Using a simple numerical approximation, we find $\Bar{\alpha} \sim 1.56$. In Fig.~\ref{fig:Tc_vs_alpha}, we collect all the above results and plot the mean-field critical temperature $T_c$ and the first-order transition temperature $T^\ast$ as a function of $\alpha$ for the two-dimensional Ising model in the $\alpha$-Rényi ensemble. The plot also includes results from our Monte Carlo data collapse for comparison, described in detail in Sec.~\ref{sec:Monte_Carlo} below. In Fig.~\ref{fig:m_vs_T_final_paper}, we plot the absolute value of the mean-field magnetization per spin $\overline{m}$ (i.e. the value of $m$ that minimizes $f_\alpha (m)$) as a function of temperature for various values of $\alpha$.

We now explore the critical exponents of the continuous phase transition regime predicted by the $\alpha$-Rényi ensemble, i.e., for $1\le \alpha \le \frac{\sqrt{13}-1}{2}$. Since the $\alpha$-Rényi mean-field free energy can be written analytically in terms of $m$ for $T\sim T_c$ (where we have $|\overline{m}|\sim0$) by performing a Taylor approximation of the logarithm (see Eq.~(\ref{eq:free_energy_6th_order}) for the $\mathcal{O}(m^6)$ expression), there must exist critical exponents that describe the behavior of the magnetization and the divergences of thermodynamic quantities such as the specific heat and magnetic susceptibility within mean-field theory. Specifically, our goal is to derive the dependence of these exponents on $\alpha$, if any. To that end, we only require the free energy to $4^\text{th}$ order in $m$, $f^{(4)}_\alpha (m)$, equivalent to Eq.~(\ref{eq:free_energy_6th_order}) less the $6^\text{th}$ order term. We find that the mean-field critical exponents $\beta$, $\alpha_{c_v}$, $\gamma$ and $\delta$ (we denote the specific heat critical exponent as $\alpha_{c_v}$) take the exact same values for the Ising model in the Rényi ensemble as they do in the Gibbs state: $\beta=1/2$, $\alpha_{c_v} = 0$, $\gamma = 1$ and $\delta = 3$ . 

While the coefficients modulating the divergences of some of the quantities of interest (e.g. $A_+$ and $A_-$, as in $\chi \sim A_+ |t|^{-\gamma}$ for $t>0$, where $t\equiv |T-T_c|/T_c$ is the reduced temperature and $\chi$ is the susceptibility) do indeed depend on the Rényi index $\alpha$, in mean-field theory, we find that the critical exponents listed above do not---a remarkable result. The last of the relevant critical exponents for this discussion is $\nu$, which is the critical exponent describing the divergence of the correlation length $\xi$ according to $\xi \sim |t|^{-\nu}$. For the Ising model in the Gibbs state, its derivation involves re-introducing fluctuations into the partition function with the derivation heavily reliant on the presence of the Gibbs state exponentials \cite{Kardar}. An attempt at following an analogous argument for the $\alpha$-Rényi ensemble presents us with the challenge of evaluating a partition function whose number of terms depends on the temperature-dependent Rényi constraint in Eq.~(\ref{eq:alphaRenyiEnsConstr}), and whose terms depend on the average energy, which appears daunting to solve for analytically for most values of $\alpha$. However, mean-field theory predicts that $\beta$, $\alpha_{c_v}$, $\gamma$ and $\delta$ not only all exist for $1\le \alpha \le \alpha_{[c\rightarrow \text{1st}]}$ but are also $\alpha$-independent, and since the divergence of thermodynamic quantities is understood in statistical physics to stem from the divergence of the correlation length and the resulting scale invariance, we claim that $\nu$ takes on the mean-field Gibbs state value of $1/2$ for all $\alpha$. 

\section{1D Ising Model: Analytical arguments} \label{sec:1D_exact_solution_final_paper}

\textit{Extremal Cases---}We now focus on the  1D Ising model given by $H = -J \sum_{i=1}^N \sigma_i \sigma_{i+1}$ with periodic boundary conditions $\sigma_{N+1}=\sigma_1$. It is known that the classical Ising model in 1D exhibits no spontaneous symmetry breaking at finite temperature in the Gibbs state. We investigate if long-range order at finite $T$ is possible in the generalized $\alpha$-Rényi ensemble and, if so, at what values of $\alpha$. We start with the limit $\beta\rightarrow\infty$. The constraint in Eq.~(\ref{eq:alphaRenyiEnsConstr}) becomes
\begin{equation}
    E_j \le \lim_{\beta\rightarrow \infty} \left[\frac{\alpha}{\beta (\alpha-1)} + \Bar{E}\right] = \Bar{E}, \nonumber.
\end{equation}
At zero temperature, all allowed microstates have energies lower than the average energy. Since 
\(\Bar{E} = \sum_{j=0}^{n_\beta-1} N_j E_j p_j^{(\alpha)}\) 
is a convex combination of the allowed \(E_j\) values, the only way to ensure \(E_j \leq \Bar{E}\) for all states is if only the ground state is allowed, where all spins are aligned. Thus, \(\Bar{E} = E_0\), where \(E_0\) is the ground state energy.

In the thermodynamic limit, higher-energy fixed points can exist at \(T=0\) (as discussed in the finite temperature section), but the fixed point that minimizes the free energy globally must be the ground state, since at \(T=0\), the free energy is simply \(\Bar{E}\). Additionally, while a finite system can equally occupy the all-up and all-down ground states, an infinite system, assuming only local fluctuations, must choose one of these configurations and thus symmetry is broken at zero temperature in this limit.

Similarly, in the $\beta = 0$ limit, we can see from the Rényi constraint that all microstates become accessible, for all $\alpha \ge 1$. The partition function evaluates to $2^N$, all probabilities equalize as $p_j^{(\alpha)} = 1/2^N$, and the magnetization vanishes. 

\textit{Finite Temperature---} In the case of the Ising model in the Gibbs state, the solution is found by evaluating the partition function analytically and using the result to derive the magnetization. Instead, we follow a different strategy and make a Peierls argument \cite{Peierls1936}. If the system starts in one of the two symmetry-broken ground states at $T=0$, and $T$ is then increased, if there is enough thermal energy to excite the system into flipping a single spin, then the minority droplet of flipped spins can grow and move until all states with two broken bonds become accessible with equal probability via local thermal fluctuations. The system can now reach the other symmetry-broken regime, making both ground states equally probable, and the magnetization vanishes. The first excited state in 1D with energy $E_1 = -JN +4J$ has two broken bonds. Once $E_1$ is "turned on", any long-range order is destroyed. Our goal now is to solve the fixed point equation Eq.~(\ref{eq:fixed_point_equation}) at finite $\beta$, and ultimately determine if higher energy fixed points $\bar{E}>E_0$ are allowed at any finite $T$.

Firstly, we note that at a given finite temperature, there may be multiple fixed points $\Bar{E}_{fp}$ that solve Eq.~(\ref{eq:fixed_point_equation}). To demonstrate this, let us assume that the system is in a state such that it can access only one of the two $E_0$ configurations.  If $\bar{E} = E_0$ is a fixed point, then $E_1$ must violate the constraint, i.e.
\begin{align}
    E_1 &> \frac{\alpha}{\beta(\alpha-1)}+ E_0, \nonumber\\
    -JN + 4J &> \frac{\alpha T}{(\alpha-1)} -JN, \nonumber
\end{align}
which produces
\begin{equation}
    T < \frac{\alpha-1}{\alpha} 4J.
\end{equation}
In other words, $\bar{E}=E_0$ is a fixed point for all $T\in\left[0,\frac{\alpha-1}{\alpha} 4J\right)$ ($T\in[0,2J)$ when $\alpha=2$), a result valid in all dimensions and in the limit $N\rightarrow\infty$. However, for $T\in\left[0,\frac{\alpha-1}{\alpha} 4J\right)$, higher energy fixed points than $\Bar{E}=E_0$ also exist in the thermodynamic limit.  Firstly, in 1D, the degeneracy of the $j^\text{th}$ energy level (with $2j$ broken bonds) is given by $N_j = 2 \times \binom{N}{2j}$ which is an $\mathcal{O}\left(N^{2j}\right)$ number. We now ask whether, given some temperature $\beta$, we can find a valid solution to Eq.~(\ref{eq:fixed_point_equation}) for $\bar{E}$ in the limit $N\rightarrow\infty$ that is a convex combination of $E_0$ and an arbitrary number of excited state energies. As an example, if we assume $E_0$ and $E_1$ are the only two accessible energies, then we have
\begin{equation}
    \frac{\sum\limits_{j=0}^1 N_j E_j \left[1-\beta\frac{\alpha-1}{\alpha}(E_j-\Bar{E})\right]^\frac{1}{\alpha-1}}{\sum\limits_{j=0}^1 N_j \left[1-\beta\frac{\alpha-1}{\alpha}(E_j-\Bar{E})\right]^\frac{1}{\alpha-1}} = \Bar{E}. \nonumber
\end{equation}
With $N_0 = 2\times \binom{N}{0} = 2$ and $N_1 = 2\times \binom{N}{2} = N(N-1)$, in the limit $N\rightarrow\infty$, the $\mathcal{O}(N^2)$ term dominates in both the numerator and denominator of the left-hand side, producing $\Bar{E} = E_1$.
Since $E_2$ was assumed a priori to be the lowest forbidden energy, then $E_2 > \frac{\alpha}{\beta(\alpha-1)} + \bar{E} = \frac{\alpha}{\beta(\alpha-1)} + E_1$, and with $E_2 = -JN + 8J$ and $E_1 = -JN+4J$, we find $T < \frac{\alpha-1}{\alpha}4J$ once again. Thus, if $E_0$ and $E_1$ are the only allowed energies, $\Bar{E} = E_1$ is a thermodynamic limit fixed point for all $T\in \left[0,\frac{\alpha-1}{\alpha} 4J\right)$. So far, that makes two fixed points in the limit $N\rightarrow\infty$ for any $T\in \left[0,\frac{\alpha-1}{\alpha} 4J\right)$: $\Bar{E}=E_0$ and $\Bar{E}=E_1$. We can continue with this line of thinking by introducing the next energy $E_2$ as an allowed energy a priori (with $E_3$ being the lowest forbidden energy), noting that $N_2$ is an $\mathcal{O}\left(N^{4}\right)$ term and that it dominates in both numerator and denominator of the left-hand side of the fixed point equation as $N\rightarrow \infty$, giving us $\Bar{E} = E_2$ as another mathematically valid solution for all $T\in \left[0,\frac{\alpha-1}{\alpha} 4J\right)$.

In this way, higher energy fixed points for any $T\in \left[0,\frac{\alpha-1}{\alpha} 4J\right)$ can be found by continuing to introduce higher energies $E_j$ as accessible states, until energies with degeneracies that have similar $N$-scaling to the maximum degeneracy of $2\binom{N}{N/2}\sim\mathcal{O}\left(2^N/\sqrt{N}\right)$ \cite{stackoverflow} are reached and multiple terms begin to survive in the fixed point equation in the limit $N\rightarrow\infty$ as opposed to the single dominant terms we have seen in the simple examples above. As a result, at each temperature, there is a maximum energy fixed point that can be found in the thermodynamic limit. We now argue that this is also true for large finite systems, and we observe numerically that this maximum grows with increasing $T$. Similarly, we argue that in the limit $N\rightarrow\infty$, this fixed point globally minimizes the free energy in any dimension $D$. The intuition behind this statement is that higher energy microstates, which are not exponentially suppressed in the Rényi ensemble, have increasing degeneracies that significantly boost the entropy, thus providing an overall lower Renyi free energy despite arising from a maximum energy fixed point. This leads to the approach we use for the attractive fixed point search in our 2D Monte Carlo simulations, the results of which we present in Sec.~\ref{sec:Monte_Carlo}. In 1D, since the maximum energy fixed point at all $T>0$ satisfies $\Bar{E}>E_0$ in the thermodynamic limit, we conclude that there is no spontaneous symmetry-breaking at finite $T$ in the 1D Ising model in the $\alpha$-Rényi ensemble.

\section{2D Ising Model: Monte Carlo}  \label{sec:Monte_Carlo}
Let us now consider the case of the two-dimensional classical Ising model in the $\alpha$-Rényi ensemble. We are interested in studying the critical behavior of the true, correlated model. A key goal of this study is to shed light onto the extent to which the Rényi ensemble reproduces the Gibbs state in light of the claims made in Refs.~\cite{Giudice2021,Giudice2024} that these two ensembles reproduce each other for local observables in the thermodynamic limit. Similarly, we want to know if any phase transition that emerges coincides with the mean-field prediction that there is a "threshold" $\alpha$ separating continuous and first-order regimes. Unlike the Onsager result for the 2D Ising model in the Gibbs state \cite{Onsager1944}, the challenges associated with evaluating the Rényi ensemble partition function in Eq.~(\ref{eq:alphaRenyiPartFunc}) make an exact solution difficult to derive, rendering the model ripe for numerical exploration.

We use the Monte Carlo (MC) method with the Metropolis algorithm to simulate the 2D Ising model in an equilibrium defined by the $\alpha$-Rényi ensemble probabilities in Eq.~(\ref{eq:alphaRenyiEnsProb}), choosing single-spin flip dynamics for simplicity. We customize the original Metropolis algorithm \cite{Metropolis1953} and define
\begin{align} \label{eq:metropolis_algorithm_Renyi}
&A_\alpha(\mu \rightarrow \nu) = \nonumber\\
&\begin{cases}
    \frac{\left[1 - \beta \frac{\alpha-1}{\alpha}(E_\nu - \Bar{E})\right]^\frac{1}{\alpha-1}}{\left[1 - \beta \frac{\alpha-1}{\alpha}(E_\mu - \Bar{E})\right]^\frac{1}{\alpha-1}}, & \text{ } E_\nu > E_\mu \text{ }\big|\big|\text{ } E_\nu \le \frac{\alpha}{\beta (\alpha-1)} + \Bar{E} \\
    1, & \text{ } E_\nu \le E_\mu \text{ }\big|\big|\text{ } E_\nu \le \frac{\alpha}{\beta (\alpha-1)} + \Bar{E} \\
    0, & \text{ } E_\nu > \frac{\alpha}{\beta (\alpha-1)} + \Bar{E},
\end{cases}
\end{align}
where $A_\alpha(\mu \rightarrow \nu)$ represents the acceptance ratio associated with a transition from the current state $\mu$ to a proposed state $\nu$ (parametrized by the Rényi index $\alpha$) and where it is assumed that the system is already in a state $\mu$ that satisfies the Rényi constraint prior to the update.

The algorithm satisfies detailed balance, but it is not always ergodic. For example, at $T\approx 0$, we have $\bar{E}\approx E_0$ with essentially only the two degenerate ground states allowed, as discussed in the previous section. A Monte Carlo simulation at that temperature can be initialized in one of those ground states, but to reach one starting from the other using local dynamics would require accessing excited states that are strictly forbidden by the Rényi constraint. This inaccessibility problem resolves when all energies are allowed, that is, when the maximal energy $E_j=2JN$ is accessible as per Eq.~(\ref{eq:alphaRenyiEnsConstr}). For a given $\alpha$, this occurs for all $\left(T,\bar{E}\right)$ pairs that satisfy
\begin{align} \label{eq:ergodicity_requirement}
    T &\ge \frac{\alpha-1}{\alpha}(2JN-\bar{E}(T))
\end{align}
with $\bar{E}$, the solution to Eq.~(\ref{eq:fixed_point_equation}), here denoted $\bar{E}(T)$ to emphasize that it is a function of $T$. While ergodicity is anticipated if the inequality is satisfied, for example at very large $T$ when $\bar{E}\sim0$, it does not necessarily break down if this is not the case. This is due to the fact that even if the maximal energy is forbidden at some temperature, a local dynamics algorithm acting on a system initialized in any of the allowed configurations may not need to access the maximal energy states to be able to reach the other allowed configurations with nonzero probability. As discussed, our single-spin flip algorithm starts off non-ergodic at $T=0$, but then as $T$ increases, it is expected to become ergodic at some $N$-dependent temperature, $T=T_\text{erg}(N)$.

Similarly, the breakdown of ergodicity at finite temperature resolves as $\alpha$ approaches $1$ and the Rényi ensemble tends to the Gibbs state (where all energies are allowed), with the right-hand side of Eq.~(\ref{eq:ergodicity_requirement}) tending to $0$, i.e., $\lim_{\alpha\rightarrow1}T_\text{erg}=0$. 
We argue that these ergodicity issues do not affect our analysis for the observables we consider in our simulations. This is similar to a Monte Carlo simulation of the Gibbs state at low temperature, where simulations of the 2D Ising model for large system sizes result in excellent approximations of the critical temperature and critical exponents. These simulations can be conducted in such a way that only one of the symmetry-broken regimes ($m>0$ or $m<0$) ends up being explored in the typical amount of Monte Carlo time for which such simulations are usually performed without affecting the determination of observables and critical exponents \cite{LandauBinder}.

In our Rényi ensemble simulations, the lack of ergodicity at low temperatures arises due to the system being unable to cross from one symmetry-broken regime into the other. Within each regime, the simulation is expected to be ergodic; in other words, if the system is in an $m>0$ ($m<0$) mode, it will be able to access all other $m>0$ ($m<0$) configurations that are not forbidden by the constraint in Eq.~(\ref{eq:alphaRenyiEnsConstr}). We expect that collecting data from only one symmetry-broken regime will be enough to characterize any phase transition that we detect in the 2D Ising model, even if the inability to collect data from the other regime is not due to lack of Monte Carlo time, but instead due to the breakdown of ergodicity. Thus, we do not expect the lack of ergodicity between $m>0$ and $m<0$ configurations in specific $(T,\alpha,N)$ parameter regimes to affect the study of the critical behavior of the model. A cluster algorithm such as the Wolff algorithm \cite{Wolff1989} may totally avoid ergodicity breakdown at all $T\ge0$, but we see no need to go beyond local dynamics for the purposes of our specific study.

\begin{figure}
\includegraphics[scale=0.57]{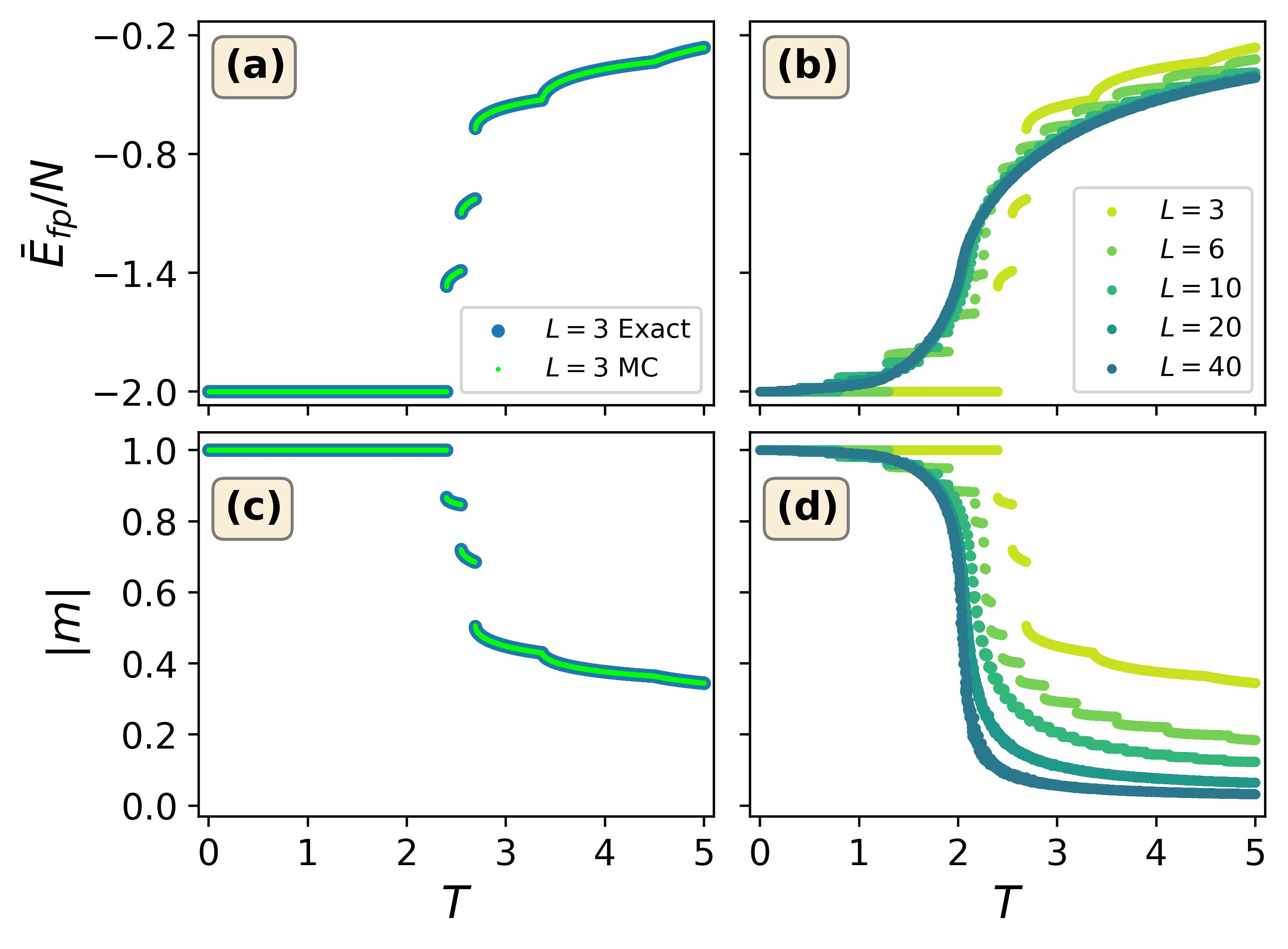}
\caption{\label{fig:epsart} \textbf{(a)}-\textbf{(b)} Average energy per spin $\bar{E}_\text{fp}/N$ (with $N=L^2$) computed using the fixed point search technique of Sec.~\ref{sec:Monte_Carlo} and \textbf{(c)}-\textbf{(d)} absolute value of the magnetization per spin $|m|$ as a function of temperature $T$ for the 2D Ising model in the $2$-Rényi ensemble. In \textbf{(a)} and \textbf{(c)}, the Monte Carlo (MC) results for a $3\times3$ system are compared with exact results. In \textbf{(b)} and \textbf{(d)}, the effects of changing system size on the MC results are shown. The small system results exhibit discontinuous jumps in $\bar{E}_\text{fp}/N$ and $|m|$ with changing $T$, hinting at the possibility of a first-order transition in the thermodynamic limit, but as $L$ increases, the curves begin to display a more continuous character, providing evidence of a continuous phase transition. For an explanation of the discontinuities, we refer the reader to the last paragraph of Sec.~\ref{sec:Monte_Carlo}. At all $T$, the MC error bars corresponding to the "minimum errors" discussed in Sec.~\ref{sec:Monte_Carlo} are smaller than the size of the data points.
}
\label{fig:alpha2.00_FourPlots}
\end{figure}

A critical issue that must be resolved in order to simulate the Rényi ensemble is the presence of the average energy $\bar{E}$ in the corresponding probabilities and by association the acceptance ratio in Eq.~(\ref{eq:metropolis_algorithm_Renyi}). At each temperature $\beta$, we must solve for $\Bar{E}$ by solving the fixed point equation Eq.~(\ref{eq:fixed_point_equation}). In App.~\ref{appendix:attractive_fixed_point}, we argue that the $2$-Rényi ensemble has an attractive fixed point for the Ising model with no external field, and we expect this to remain true for all $\alpha\ge 1$. 

For each temperature $T$ and Renyi index $\alpha$, we start by pre-selecting an initial value of $\Bar{E}$, defined as $\Bar{E}^{(0)}$, which allows the Rényi acceptance ratio Eq.~(\ref{eq:metropolis_algorithm_Renyi}) to be fully characterized. We use this ratio to perform a full Monte Carlo simulation of the 2D Ising model and extract a new estimate of the average energy $\Bar{E}^{(1)}$ using importance sampling and the binning technique~\cite{BarkemaNewmanBook, BeccaSorellaBook}. The attractive nature of the fixed point means that, unless $\Bar{E}^{(0)}$ is true fixed point $\Bar{E}_\text{fp}$, $\Bar{E}^{(1)}$ should be closer to $\Bar{E}_\text{fp}$ than $\Bar{E}^{(0)}$, barring Monte Carlo errors in the estimation of the average energies. Next, we take $\Bar{E}^{(1)}$, plug it into Eq.~(\ref{eq:metropolis_algorithm_Renyi}) to form a new acceptance ratio, and repeat the process to extract $\Bar{E}^{(2)}$ at the new equilibrium. We continue in this vein until we have found some $\Bar{E}^{(k)} \approx \Bar{E}_\text{fp}$ after $k$ Monte Carlo simulations. 
 
The fixed point search is defined by the recursion
\begin{align} \label{eq:fixed_point_search_recursion_MC}
    \begin{split}
        &... \\
        g_\alpha \left(\bar{E}^{(i-1)}\right) &\approx \bar{E}^{i} \\
        g_\alpha \left(\bar{E}^{(i)}\right) &\approx \bar{E}^{i+1} \\
        &...,
    \end{split}
\end{align}
where $g_\alpha \equiv \Tr(\rho^{(\alpha)} H)$. Here, there is Monte Carlo error involved in the estimation of $g_\alpha \left(\bar{E}^{(i)}\right)$ for every $i$ in the iteration. This noise is propagated through the recursion as the simulation searches for the fixed point, but we find that, at each step in the recursion, if the Monte Carlo time is large enough and an accurate estimation process based on the binning technique is used, this noise has little effect when it comes to moving in the general direction of the fixed point and ultimately extracting a reasonable estimate for $\Bar{E}_\text{fp}$. 

To identify the fixed point, we choose to define a new hyperparameter $N_\text{osc}$ that counts the number of times the fixed point search "oscillates". In other words, once the general vicinity of the fixed point has been approximately found, its attractive nature means that continuing the recursion should make the Monte Carlo estimate for $\bar{E}$ oscillate about some average value that is very close to the true $\bar{E}_\text{fp}$, and we quantify this oscillation by counting the number of times $\left(\Bar{E}^{(i+1)} - \Bar{E}^{(i)}\right)$ changes sign from one iteration to the next, defining $N_\text{osc}$ as precisely this number. In practice, we find that as long as $N_\text{osc}$ is large enough, changing its value does not significantly affect the final results for the Monte Carlo averages and data collapse.

At each temperature $T$, there may be more than one fixed point. In Sec.~\ref{sec:1D_exact_solution_final_paper}, we showed that in the 1D model, the Rényi ensemble can generate a large number of fixed points at each temperature in the thermodynamic limit. While the analysis to prove this in the 2D Ising model would be more involved, Monte Carlo simulations with fixed point searches initialized at different values of $\bar{E}$ (i.e. different $\bar{E}^{(0)}$) provide evidence for the existence of multiple fixed points at most $T$ (results not shown), for the finite system sizes that we choose to study. 

Following the discussion in Sec.~\ref{sec:1D_exact_solution_final_paper}, we adopt the "maximum energy" fixed point approach at each temperature. We begin at $T = 0.001$, where we expect the maximum energy fixed point to be near the ground-state energy $E_0 = -2JN$. To ensure we capture the maximum energy fixed point, we initiate the search from an energy above $E_0$, setting the initial average energy to $\bar{E}^{(0)} = E_0 + \Delta$, where $\Delta$  is a sufficiently large offset.

The search terminates once $N_\text{osc} = 30$ oscillations are detected for $\bar{E}^{(i)}$. Letting $k$ be the number of iterations of Eq.~(\ref{eq:fixed_point_search_recursion_MC}) needed to reach $N_\text{osc} = 30$, and $\bar{E}^{(k)}$ the average energy at the $k^\text{th}$ iteration, we then perform one final simulation with $\bar{E} = \bar{E}^{(k)}$ in Eq.~(\ref{eq:metropolis_algorithm_Renyi}), extending the MC time significantly to obtain the final average energy estimate, which we denote as the maximum energy fixed point $\bar{E}_\text{fp} \approx \bar{E}^{(k+1)}$. This final run also provides the estimate for the magnetization $|m|$ and its associated error bar. Throughout all simulations, the all-down ground state is chosen as the initial configuration.

With the $T=0.001$ simulation now complete, we seek results for $T\in \left[0.001,5.000\right]$ in increments of $dT=0.001$. We increment $T$ as $T\rightarrow T+dT$, and, at every subsequent temperature, we embark on an annealing strategy for the fixed point search defined by
\begin{align} \label{eq:annealing_relation}
\begin{split}
    \bar{E}^{(0)}(T+dT) & = \bar{E}_\text{fp}(T) + \Delta_E \\ &\approx \bar{E}^{(k+1)}(T) + \Delta_E. 
\end{split}
\end{align}
In other words, for each temperature $T+dT$ we set the initial average energy used in the search for the elusive fixed point equal to the fixed point estimate from the previous temperature $T$ plus some $\Delta_E$ that must be large enough to ensure we are conducting the next search from above. We note that $k$, the number of simulations required to reach $N_\text{osc}=30$, is temperature-dependent.

\begin{figure}
\includegraphics[scale=0.57]{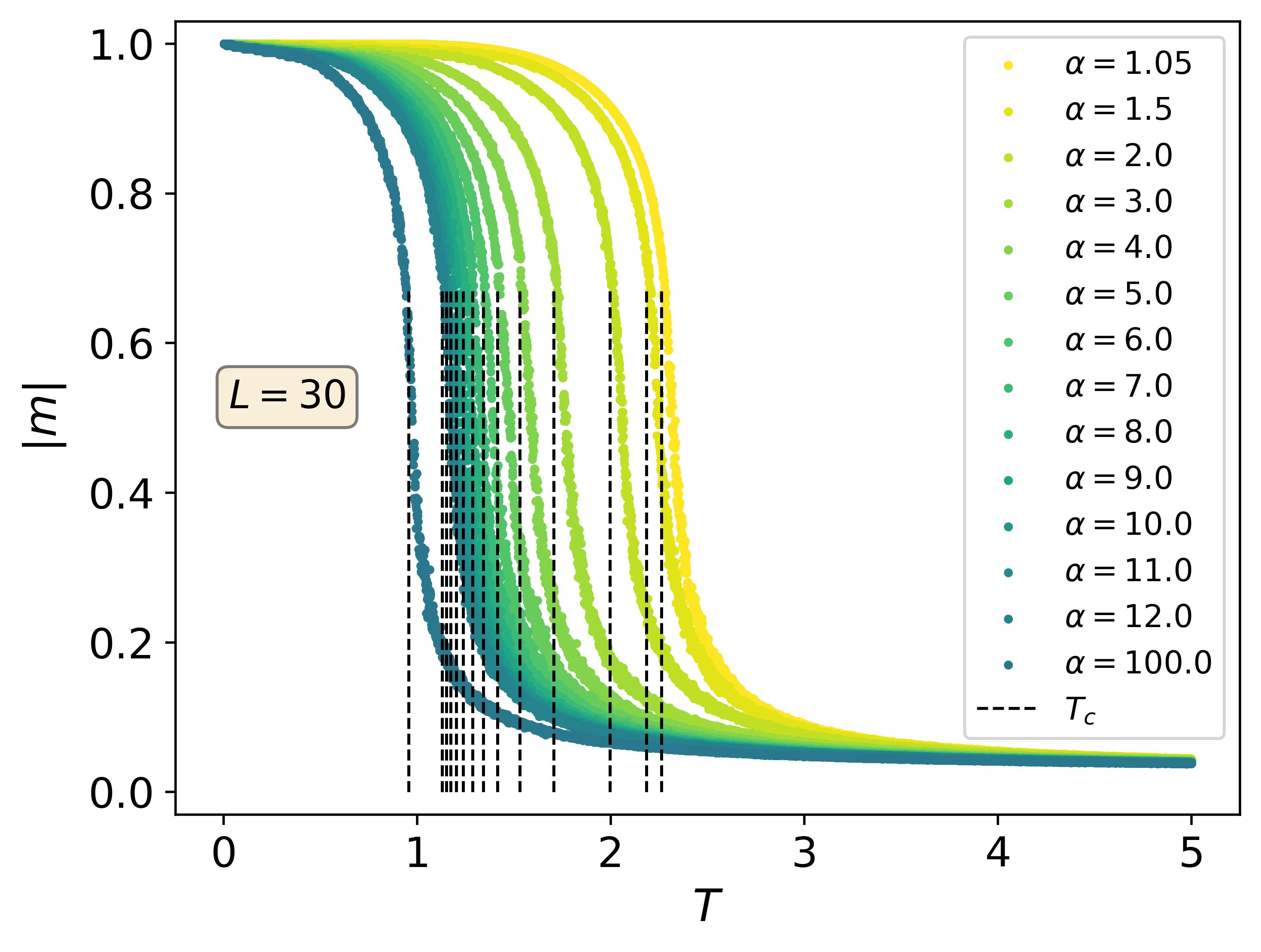}
\caption{\label{fig:epsart} Absolute value of the Monte Carlo magnetization per spin $|m|$ as a function of temperature $T$ for the 2D Ising model in the $\alpha$-Rényi ensemble at $L=30$, for a variety of values of $\alpha$. For each curve, the corresponding estimate of $T_c$ as extracted by collapsing the MC data using Eq.~(\ref{eq:RG_scaling_function_Gibbs_state}) is shown. The curves are relatively continuous in character, hinting at a continuous phase transition for all $\alpha$. The shape of the curves through the transition appears to be independent of $\alpha$, but the estimates of $T_c$ exhibit a strong $\alpha$-dependence. The error bars are smaller than the size of the data points, but these are "minimum errors" as previously described.
}
\label{fig:m_vs_T_MonteCarlo_L30_allAlpha}
\end{figure}

In Fig.~\ref{fig:alpha2.00_FourPlots}, we plot Monte Carlo results at $\alpha=2$ for the average energy per spin $\bar{E}_\text{fp}/N$ and absolute value of magnetization $|m|$ as a function of $T$ for various system sizes of interest (see App.~\ref{appendix:MC_stats} for details on autocorrelation time and thermalization). The near-perfect overlap between Monte Carlo and exact results in the $3\times3$ case highlights the strength of the MC approach and the accuracy of the annealing method we use for the fixed point search. We note that the exact results were computed to high precision by leveraging the attractive nature of the fixed point as well, with each search starting from above at exactly $\bar{E}^{(0)}=0$. For the $3\times3$ results, $\bar{E}_\text{fp}/N$ and $|m|$ exhibit discontinuous jumps at various temperatures, hinting at the possibility of a first-order transition in the thermodynamic limit. However, the results at larger $N$ provide evidence for the presence of a continuous transition at $\alpha=2$, with the curves becoming more and more continuous with increasing $N$, and tending to the shapes that are typically observed in the $\alpha\rightarrow 1$ case for the 2D classical Ising model \cite{BarkemaNewmanBook}.

Turning to Monte Carlo error, the error bars in Fig.~\ref{fig:alpha2.00_FourPlots} are nominally smaller than the size of the data points, but these errors must be termed "minimum errors", because the fixed point $\bar{E}^{(k+1)}\approx\bar{E}_\text{fp}$ found at each temperature after the final Monte Carlo simulation is an approximation and not exact, with $\big|\bar{E}^{(k+1)} - \bar{E}^{(k)}\big|$ small but nonzero. We do not attempt to quantify the error beyond computing the errors of the final averages as per the procedure outlined in Ref.~\cite{BeccaSorellaBook}. 

\begin{figure}
\includegraphics[scale=0.54]{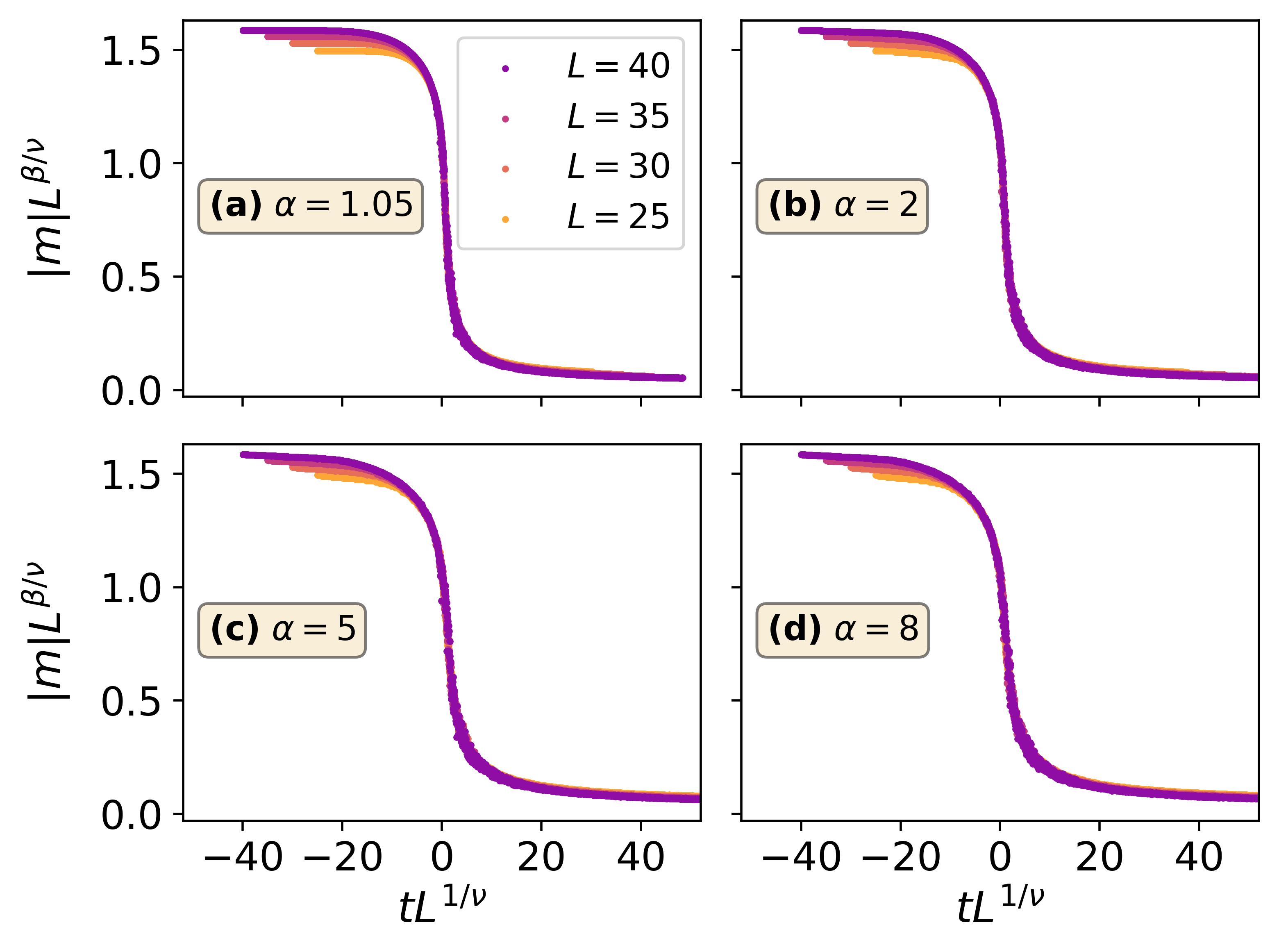}
\setlength{\textfloatsep}{1pt }
\setlength{\abovecaptionskip}{0.8pt} 
\setlength{\belowcaptionskip}{4pt} 
\caption{\label{fig:epsart} Results for the collapse of the Monte Carlo magnetization data for the 2D Ising model, for four different values of $\alpha$. The data collapses quite well for all $\alpha$ in the vicinity of $t\sim 0$. The collapse was performed as follows: $\beta$ and $\nu$ are fixed to the Gibbs state ($\alpha\rightarrow1$) critical exponent values for the 2D Ising model, and $T_c$ is tuned using the polynomial fit technique described in Sec.~\ref{sec:Monte_Carlo}, with a $25^\text{th}$ order polynomial used to fit $5\%$ of the data either side of $t=0$.
}
\label{fig:dataCollapse_FourPlots}
\end{figure}

\begin{figure*}
\includegraphics[scale=0.8]{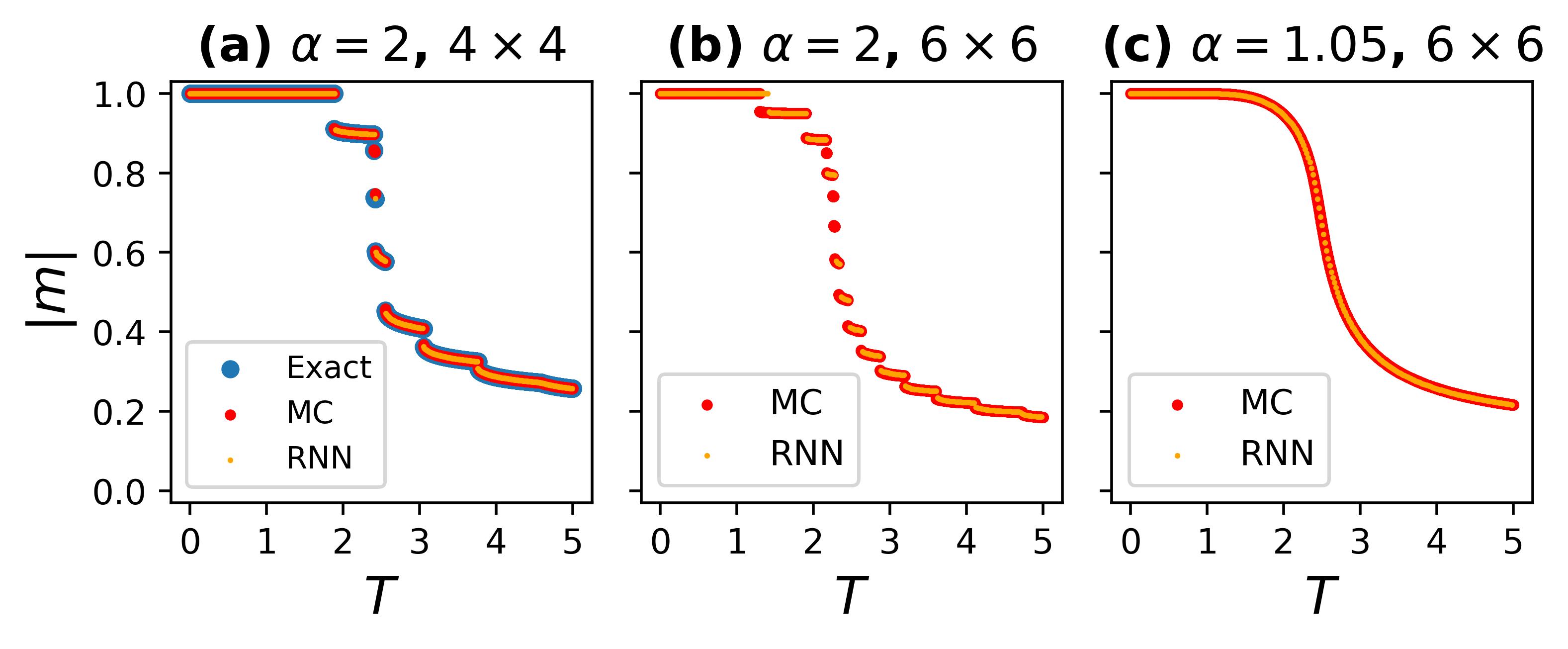}
\vspace{0pt}
\caption{\label{fig:epsart} Comparison between RNN, Monte Carlo and exact results for the absolute value of the magnetization per spin $|m|$ of the 2D Ising model in the $\alpha$-Rényi ensemble as a function of temperature $T$, for three combinations of system size and Rényi index. The exact and Monte Carlo results are based on the maximum energy fixed point approach of Sec.~\ref{sec:1D_exact_solution_final_paper}. A single-layer RNN was used with $50$ memory units \cite{MohamedRNN2020}, $500$ samples for gradient estimation, $2\times10^4$ training steps and $10^6$ ($2\times 10^5$) samples for the final observable estimation for the $4\times4$ ($6\times6$) system. For the Monte Carlo and exact results, the data points are produced in increments of $dT=0.001$, while $dT=0.01$ $(0.02)$ for the $4\times4$ ($6\times6$) RNN results. Overall, the RNN performs strongly for system sizes up to $6\times6$, despite some minor discrepancies near certain discontinuities when the RNN appears to find lower energy fixed points as per the analysis in Sec.~\ref{sec:RNN}. We should note that some of the discrepancies near the discontinuities in panels (\textbf{a}) and (\textbf{b}) may be partially due to there being less RNN data points compared to the Monte Carlo output ($dT = 0.02$ vs $dT=0.001$). At $\alpha=1.05$ in panel \textbf{(c)}, $|m|$ becomes "Gibbs"-like in its continuous nature when compared to the $\alpha=2$ results in panel \textbf{(b)}.} 
\label{fig:mvsT_3Plot}
\end{figure*}

In Fig.~\ref{fig:m_vs_T_MonteCarlo_L30_allAlpha}, we plot the Monte Carlo results for the magnetization $|m|$ as a function of $T$ for an $L=30$ Ising model in 2D at various values of $\alpha$. We find that for each $\alpha$ probed, the shape of the curve remains effectively the same, generally continuous in character, indicating a continuous phase transition in the thermodynamic limit. However, the position of the critical temperature $T_c$, which we estimate using the data collapse technique described below, shifts with increasing $\alpha$. The continuous nature of the curves becomes even more apparent as $L$ is increased beyond $30$ (not shown). 

To extract $T_c$, we take inspiration from renormalization group (RG) scaling theory for continuous phase transitions in classical models in the Gibbs state, elucidated in full detail in Refs.~\cite{ShankarBook,GoldenfeldBook,NishimoriOrtizBook,BarkemaNewmanPaper1996,RiegerYoung1993}. Specifically, we make use of the following scaling function:
\begin{equation} \label{eq:RG_scaling_function_Gibbs_state}
    |m|L^{\beta/\nu} = f\left(tL^{1/\nu}\right).
\end{equation}
Here, $\beta$ is the magnetization critical exponent, i.e.,  as $T\rightarrow T_c^{-}$, $m \sim (T_c-T)^\beta$, $t\equiv (T - T_c) / T_c$ is the reduced temperature, and $\nu$ is the critical exponent that characterizes the divergence of the correlation length as the critical point is approached, i.e.  $\xi \sim |t|^{-\nu}$. Eq.~(\ref{eq:RG_scaling_function_Gibbs_state}) tells us that given access to high quality Gibbs state ($\alpha\rightarrow1$) data for $|m|$ vs $T$ for all values of $L$, all the data points collapse onto a single $|m| L^{\beta/\nu}$ vs $tL^{1/\nu}$ curve.

Eq.~(\ref{eq:RG_scaling_function_Gibbs_state}) is only valid in the vicinity of the critical point (i.e. near $t=0$), because the analysis that produces it is a single RG step performed under the assumption that $|t|=|(T-T_c)/T_c|\ll 1$ \cite{NishimoriOrtizBook,GoldenfeldBook}. This means that in theory, only data points corresponding to temperatures near $T=T_c$ should form part of this collapse. Although 
Eq.~(\ref{eq:RG_scaling_function_Gibbs_state}) applies to the Ising model in the Gibbs state, we note that the $|m|$ vs $T$ curves in Fig.~\ref{fig:m_vs_T_MonteCarlo_L30_allAlpha} do not seem to change shape significantly as $\alpha$ departs from $1$. This suggests that the collapse may also apply to all $\alpha > 1$ so long as the system is large enough and the curves are "continuous enough" through the transition. To obtain a collapse, we must tune $T_c$, $\beta$ and $\nu$. We recall that the Gibbs state values for the 2D Ising model as derived by Onsager are given by $T_c = 2J/\left[\log\left(1+\sqrt{2}\right)\right] \sim 2.269J$, $\beta = 1/8$ and $\nu = 1$.

In Fig.~\ref{fig:dataCollapse_FourPlots}, the results for the data collapse of the magnetization data associated with four relatively large values of $L$ are shown, for four different values of $\alpha$. The collapse is performed by fixing the critical exponents to the Gibbs state values $\beta = 1/8$ and $\nu = 1$, i.e., we only  tune $T_c$ through the following procedure. At each value of $\alpha$, we select a range of $T_c$ values to test within an interval that contains the approximate critical temperature as estimated from the raw magnetization data, and perform a grid search for the value of $T_c$ in this interval that minimizes the distance between the $\left(tL^{1/\nu},mL^{\beta/\nu}\right)$ data points and a polynomial fit that includes only a certain percentage of the data points either side of $t=0$, since the collapse is only supposed to apply in the vicinity of $T\sim T_c$. A simple average for $T_c$ and rudimentary error bars are computed by varying the polynomial degree (testing degrees of 15, 20 and 25) and the specific percentage of data points above and below $t=0$ that are used in the fit ($5\%, 7.5\% \text{ and } 10\%$). We keep the percentages relatively low to focus on collapsing the data in the vicinity of $t=0$ only.

As is clear from Fig.~\ref{fig:dataCollapse_FourPlots}, the data collapses well with this approach, especially near $t\sim0$. We note that we did try varying the critical exponents beyond the Onsager values, using both a grid search and other optimization tools. However, to extract numerical estimates of the critical exponents with relatively small error bars would require performing Monte Carlo simulations of the Rényi ensemble at system sizes for which the fixed point search becomes computationally intractable, such as $L\sim250$ or greater \cite{HaradaBayesianCollapse}. Moreover, our numerical results, encapsulated by Fig.~\ref{fig:dataCollapse_FourPlots} and the shifted $|m|$ vs $T$ curves in App.~\ref{appendix:m_vs_T_overlap}, are consistent with the critical exponents taking the Onsager values $\beta=1/8$ and $\nu=1$ for all $\alpha \ge 1$, at least for the system sizes that we are able to simulate.

Returning to Fig.~\ref{fig:Tc_vs_alpha}, the results of this procedure showcase $T_c$ as a function of $\alpha$ as compared to both the mean-field results in 2D and the Onsager critical temperature ($\alpha \rightarrow 1$). As $\alpha$ tends to 1 from above, $T_c$ increases and approaches $\sim 2.269J$ at a decreasing rate. In the other direction, $T_c$ initially decreases at a decreasing rate as $\alpha$ grows beyond $1$, but the data points quickly revert to a decrease at an increasing rate at larger values of $\alpha$. It appears as if $T_c$ might tend to an asymptote near $T_c \sim 1$ in the limit $\alpha \rightarrow \infty$, but we have so far been unable to find an analytical argument supporting this claim. The $T_c$ values we extract seem to be a good fit when plotted against the large-$L$ magnetization data, as shown in Fig.~\ref{fig:m_vs_T_MonteCarlo_L30_allAlpha}. Additional results supporting the existence of critical behavior in the system within the Renyi ensemble are presented in App.~\ref{appendix:Chi_Binder} where we consider the behavior of the magnetic susceptibility.

Our Monte Carlo simulations indicate a continuous phase transition at a finite temperature in the 2D Ising model within the $\alpha$-Rényi ensemble for all $\alpha \ge 1$. Based on our numerical and mean-field calculations, we argue that the critical exponents remain unchanged regardless of the value of $\alpha$. While this claim requires further investigation, both analytical and numerical, further evidence supporting it can be found in App.~\ref{appendix:m_vs_T_overlap}.  
However, despite the magnetization curves exhibiting relatively good overlap for all $\alpha$ at both low ($T\sim0$) and high ($T\gtrsim 4$) temperatures (see Figs.~\ref{fig:Tc_vs_alpha} and \ref{fig:m_vs_T_MonteCarlo_L30_allAlpha}), the strong dependence of the local observable $|m|$ and the critical temperature $T_c$  on $\alpha$ prevents us from concluding that the Rényi ensemble is locally equivalent to the Gibbs state in the thermodynamic limit near the critical point. This finding contradicts the assertion in Ref.~\cite{Giudice2021} suggesting that the two ensembles are locally indistinguishable. 

It is in a way remarkable that in mean-field theory, the predicted transition is only continuous below a threshold value of $\alpha_{[c\rightarrow 1^\text{st}]} = \frac{\sqrt{13}-1}{2}\sim 1.303$.  In the Rényi ensemble, the constraint in Eq.~(\ref{eq:alphaRenyiEnsConstr}) is such that at a given value of $\alpha$, as temperatures are increased, more and more higher energy states are made accessible to the system discontinuously, each suddenly "turning on" at some temperature $T$. Now each state has a different magnetization $m$, and in the true model, consecutive eigenstates can be separated by a single spin flip, producing a small gap in energy and magnetization between such states. In the thermodynamic limit, this gap vanishes when considering quantities on a "per spin" basis, and so, as new states become accessible with increasing $T$, their emergence into phase space occurs continuously in this "per spin" context. This is essentially what takes place in our Monte Carlo simulations for all $\alpha$ as $N$ increases. The difference in mean-field is that the field at each site takes on the same value---that of the mean-field order parameter, $m$. Thus, the per-spin gaps in energy and magnetization between consecutive eigenstates do not vanish as $N\rightarrow\infty$, implying first-order behavior at a prospective transition, assuming $\alpha$ is large enough. If $\alpha$ is small, the right-hand side of Eq.~(\ref{eq:alphaRenyiEnsConstr}) is such that most states become accessible at all $T$, making continuous changes more likely. This is why mean-field theory predicts the existence of both continuous and first-order regimes.

\section{2D Ising Model: Recurrent Neural Networks} \label{sec:RNN}

Having shown that Monte Carlo methods can successfully simulate the $\alpha$-Rényi ensemble, we now turn to variational Monte Carlo (VMC)\cite{SolvingStatMechUsingNNs2019}. We take inspiration from Ref.~\cite{Giudice2021,Giudice2024} where the authors developed tensor network and RBM ansätze to variationally simulate quantum spin models in the the $2$-Rényi ensemble at finite temperature. Specifically, we leverage the recurrent neural network (RNN) approach of Refs.~\cite{MohamedRNN2020,MohamedTopologicalOrder2023,MohamedSupplementingRNN2021} and apply it to the study of the 2D classical Ising model in our ensemble of interest. The cost function we wish to minimize is the Rényi free energy Eq.~(\ref{eq:alphaRenyiFreeEnergy}), which in the case of a classical spin model can be rewritten as
\begin{equation} \label{eq:alphaRenyiFreeEnergy_spinVersion}
    F_{\alpha} = \sum_{\{\boldsymbol{\sigma}\}} P(\boldsymbol{\sigma}) E(\boldsymbol{\sigma}) - \frac{T}{1-\alpha} \log\left[\sum_{\{\boldsymbol{\sigma}\}} P(\boldsymbol{\sigma})P(\boldsymbol{\sigma})^{\alpha-1}\right],
\end{equation}
where $P(\boldsymbol{\sigma})$ is the probability associated with a configuration $\boldsymbol{\sigma}$ and $E(\boldsymbol{\sigma})$ is the corresponding energy. If $P(\boldsymbol{\sigma})$ is parameterized by the variational parameters $\{\lambda\}$ as $P(\boldsymbol{\sigma})\rightarrow P_\lambda(\boldsymbol{\sigma})$ then the gradients of Eq.~(\ref{eq:alphaRenyiFreeEnergy_spinVersion}) are given by

\begin{figure*}
\includegraphics[scale=1.0]{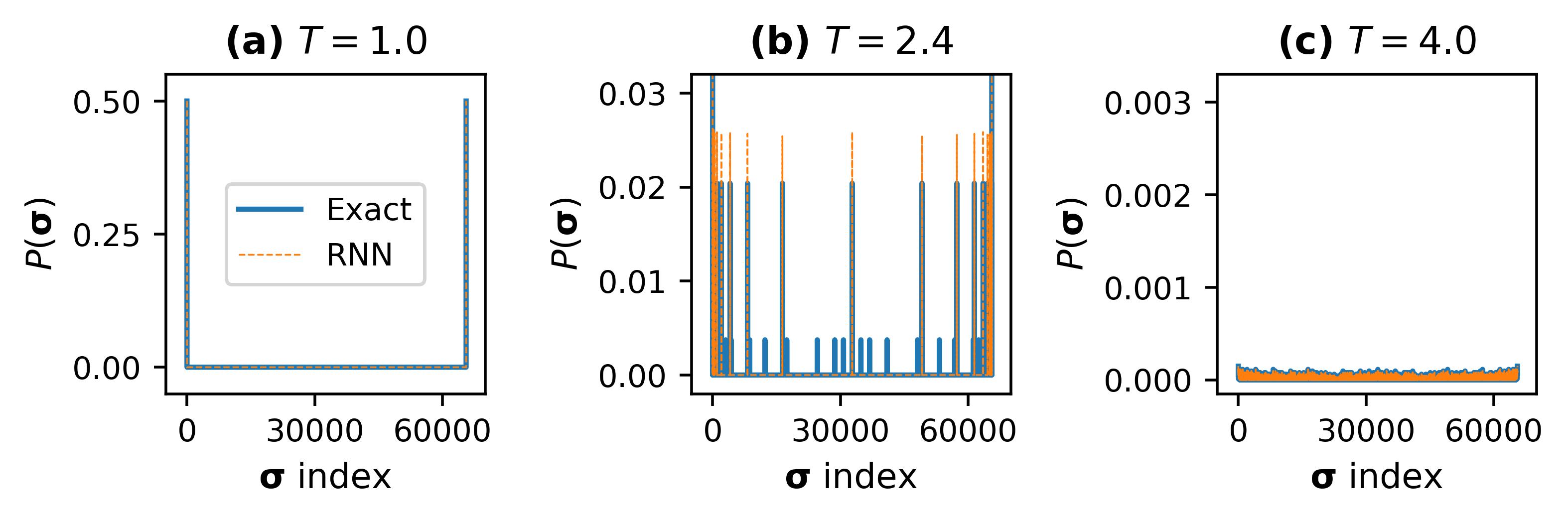}
\caption{\label{fig:epsart} Comparison between the exact 2-Rényi ensemble probability distribution (corresponding to the maximum energy fixed point) and the distribution of the trained RNN for a $4\times4$ system corresponding to the 2D Ising model results in Fig.~\ref{fig:mvsT_3Plot}(a) at \textbf{(a)} $T = 1.0$, \textbf{(b)} $T = 2.4$ and \textbf{(c)} $T = 4.0$. At all temperatures, the RNN finds a mixture of symmetric positive and negative magnetization states. For example, it is clear that at $T=1.0$ and $T=2.4$, both ferromagnetic ground states (corresponding to the first and last $\boldsymbol{\sigma}$ indices in all subplots) are generated with equal probability. The exact results are computed in such a way to also display this mixture. At $T=2.4$, the RNN finds a fixed point that is both lower in energy and free energy compared to the maximum $\Bar{E}_\text{fp}$, showing that the maximum energy approach does not always result in the global free energy minimum for small systems near energy discontinuities, as per the discussion in Sec.~\ref{sec:RNN}.
}
\label{fig:PvsSigma}
\end{figure*}

\begin{align} \label{eq:RNN_gradients_without_gradients}
\begin{split}
    &\partial_\lambda F_\alpha = \sum_{\{\boldsymbol{\sigma}\}} P_\lambda(\boldsymbol{\sigma}) \left[\partial_\lambda \log P_\lambda(\boldsymbol{\sigma})\right] E(\boldsymbol{\sigma}) \\ & - \frac{\alpha T / (1-\alpha)}{\sum\limits_{\{\boldsymbol{\sigma}\}} P_\lambda(\boldsymbol{\sigma}) P_\lambda(\boldsymbol{\sigma})^{\alpha-1}} \, \sum_{\{\boldsymbol{\sigma}\}} \biggr[P_\lambda(\boldsymbol{\sigma}) \left[\partial_\lambda \log P_\lambda(\boldsymbol{\sigma})\right] \\ &P_\lambda(\boldsymbol{\sigma})^{\alpha-1} \biggr].
\end{split}
\end{align}

For large systems, the sum $\sum_{\{\boldsymbol{\sigma}\}}$ cannot be performed exactly---instead it must be evaluated by drawing independent samples $\{\boldsymbol{\sigma^{(i)}}\}$ from $P_\lambda(\boldsymbol{\sigma})$, rendering the evaluation of the gradients stochastic. Independent sample generation is achieved directly by exploiting the autoregressive nature of the RNN, which avoids the autocorrelation issues that plague Markov-chain Monte Carlo approaches; for more detail on this, we refer the interested reader to Ref.~\cite{MohamedRNN2020}. The presence of what we call the "generalized purity" $\sum_{\{\boldsymbol{\sigma}\}} P_\lambda(\boldsymbol{\sigma}) P_\lambda(\boldsymbol{\sigma})^{\alpha-1}$ in the denominator of the second term of Eq.~(\ref{eq:RNN_gradients_without_gradients}), a quantity that tends to zero rapidly with increasing temperature and increasing system size and one that must be approximated stochastically, complicates the VMC process due to the resulting large variance of the gradient estimate. To mitigate this issue, we use a variance reduction technique proposed in Refs.~\cite{SolvingStatMechUsingNNs2019,DeepLearningBook,ThirdVarianceReductionReference2014}, which modifies the gradients as
\begin{align} \label{eq:RNN_gradients_with_gradients}
\begin{split}
    &\partial_\lambda F_\alpha = \sum_{\{\boldsymbol{\sigma}\}} P_\lambda(\boldsymbol{\sigma}) \left[\partial_\lambda \log P_\lambda(\boldsymbol{\sigma})\right] \left[E(\boldsymbol{\sigma}) - E\right] \\ & - \frac{\alpha T / (1-\alpha)}{\sum\limits_{\{\boldsymbol{\sigma}\}} P_\lambda(\boldsymbol{\sigma}) P_\lambda(\boldsymbol{\sigma})^{\alpha-1}} \sum_{\{\boldsymbol{\sigma}\}} \biggr[P_\lambda(\boldsymbol{\sigma}) \left[\partial_\lambda \log P_\lambda(\boldsymbol{\sigma})\right] \\ &\bigg(P_\lambda(\boldsymbol{\sigma})^{\alpha-1} - \sum_{\{\boldsymbol{\sigma}\}} P_\lambda(\boldsymbol{\sigma}) P_\lambda(\boldsymbol{\sigma})^{\alpha-1}\bigg)\biggr].
\end{split}
\end{align}
It can be shown that the new terms in Eq.~(\ref{eq:RNN_gradients_with_gradients}) do not bias the gradient estimates \cite{MohamedRNN2020}. The base parameterization we select for $P_\lambda(\boldsymbol{\sigma})$ is an RNN with a two-dimensional tensorized gated recurrent unit cell (2D GRU) of Ref.~\cite{MohamedSupplementingRNN2021}, which we couple to the periodic RNN structure introduced in Ref.~\cite{MohamedTopologicalOrder2023} with a two-dimensional sampling path. We use the Adam optimizer of Ref.~\cite{AdamOptRef2018} to update the parameters.

In Fig.~\ref{fig:mvsT_3Plot}, the RNN magnetization results for three different combinations of the Rényi index $\alpha$ and system size are compared with relevant exact and Monte Carlo results, both computed using the maximum energy fixed point approach of Sec.~\ref{sec:1D_exact_solution_final_paper}. To produce the RNN results, at each temperature, a single-layer RNN was used with selected hyperparameters (see Fig.~\ref{fig:mvsT_3Plot} caption). We anneal from $T=6$, where an RNN initialized with weights drawn from a Gaussian distribution is optimized, after which we decrement $T$ by $dT=0.01$ or $dT=0.02$ (corresponding to the $4\times4$ and $6\times6$ results respectively) and initialize the RNN at each subsequent temperature using the trained RNN from the previous temperature. The RNN performs strongly for the system sizes shown, with the $6\times6$ results in particular giving reason for optimism. Beyond $6\times6$, the gradient variance issues discussed above become more prominent and the approach struggles to converge.

At $\alpha=2$, minor discrepancies emerge at intermediate temperatures. In Fig.~\ref{fig:PvsSigma} we plot the exact $2$-Rényi ensemble probability distribution associated with the maximum energy fixed point for the $4\times4$ 2D Ising model of Fig.~\ref{fig:mvsT_3Plot}(a) and compare it with the corresponding RNN prediction in three different temperature regimes. The results display good agreement in the high and low temperature regimes, but at the specific intermediate temperature shown ($T=2.4$), the RNN finds a lower energy, less entropic fixed point, one that is shown to be lower in free energy ($F$ can be calculated exactly for small systems). While in the thermodynamic limit we expect the maximum energy fixed point to produce the global free energy minimum, on smaller systems (such as $4\times4$ here), the free energy may be minimized by a lower energy fixed point in the vicinity of discontinuities. This might explain the slight discrepancies between the $6\times6$ RNN and Monte Carlo results near intermediate temperature discontinuities in Fig.~\ref{fig:mvsT_3Plot}(b), despite the curves overlapping well overall. 

We expect that as system size increases, this effect becomes less pronounced due to the increasingly continuous nature of the curves, until at some large enough $L$, the free energy minimum is likely produced by the maximum energy fixed point at all $T$, justifying our approach to the fixed point search. Similarly, as $\alpha\rightarrow 1$, the magnetization becomes more continuous and the RNN and Monte Carlo results at $6\times6$ (see Fig.~\ref{fig:mvsT_3Plot}(c)) generate near-perfect overlap. We note that some of the discrepancies near discontinuities in Fig.~\ref{fig:mvsT_3Plot}(b) may also be partially due to the Monte Carlo results run on a denser grid than the RNN output ($dT=0.001$ for MC compared to $dT=0.02$ for the RNN). The RNN's success in finding the equilibrium fixed point at $T=2.4$ in Fig.~\ref{fig:mvsT_3Plot}(a) speaks to its expressive power as a variational ansatz \cite{Raghu2017}.

As Fig.~\ref{fig:PvsSigma} shows, in the low-temperature ferromagnetic phase, the RNN generates symmetric positive and negative magnetization configurations with the same probability, which contrasts with some of our low-temperature Monte Carlo simulations that are confined to a series of symmetry-broken configurations. This characteristic of the RNN approach proves advantageous, as it helps to justify the non-ergodic nature of our Monte Carlo method. Unlike Monte Carlo simulations that may fail to sample all possible modes, the RNN does not exhibit this bias, while still producing $|m|$ vs $T$  results that are generally consistent with those obtained via Monte Carlo.

All in all, the RNN produces promising results, but our approach is susceptible to difficulties at larger system sizes, when we expect the issue of having to estimate the generalized purity in the denominator of the gradient to have an effect on convergence. And while this approach works well for system sizes up to $6\times6$ in the 2D classical Ising model, we expect further challenges to arise when applying this approach to quantum spin models. This emphasizes the importance of the work of Ref.~\cite{Giudice2024}, where a method for gradient estimation that avoids vanishing denominators is introduced and applied to the quantum Ising model.

\section{Conclusion \& Outlook} \label{sec:Conclusion}
We have developed various techniques to study the generalized $\alpha$-Rényi ensemble thermal state approximation of the classical Ising model. First we analyzed the model at the mean-field level, and found that there is a threshold value of the Rényi index $\alpha_{c\rightarrow 1^\text{st}}=\left(\sqrt{13}-1\right)/2\sim 1.303$ separating continuous and first-order phase transition regimes. We proceeded to present an analytical argument as to why there is no finite-temperature symmetry-breaking phase transition in 1D for all values of the Rényi index $\alpha$. For 2D, we developed a Monte Carlo technique that targets the Rényi ensemble distribution by leveraging an attractive fixed point, and concluded that the true phase transition of the fully correlated model is continuous for all $\alpha \ge 1$. We argued that the Monte Carlo results, combined with the mean-field predictions, provide evidence that the critical exponents associated with this transition are independent of $\alpha$. However, the predicted critical temperature as extracted from a data collapse of the magnetization curves is strongly $\alpha$-dependent. While our numerical simulations away from the critical point at very high ($T \gtrsim 4$) and very low temperatures ($T\sim0$) support the arguments in Refs.~\cite{Giudice2021,Giudice2024} that the Gibbs state and the Rényi ensemble become locally indistinguishable in the thermodynamic limit, our results near the critical point suggest that the Rényi ensemble predictions can differ from the Gibbs ensemble even for local observables.

Turning to recurrent neural networks (RNNs), we presented variational Monte Carlo results for the simulation of the 2D classical Ising model in the Rényi ensemble, finding that the RNN of Refs.~\cite{MohamedRNN2020,MohamedTopologicalOrder2023,MohamedSupplementingRNN2021} performs strongly as a variational ansatz for system sizes up to $6\times6$. For larger systems, our approach to the optimization of the $\alpha$-Rényi free energy suffers from difficulties involved in estimating a vanishing purity in the denominator of the entropy gradient term. For this, we pay tribute to the work of Ref.~\cite{Giudice2024}, which found a way to use an RBM ansatz to simulate the 2D quantum Ising model in the 2-Rényi ensemble while avoiding vanishing denominators in the gradient.

The iterative Monte Carlo approach we developed has broader applicability beyond the Rényi ensemble and the Ising model. It can be used to study other classical Hamiltonians and extract their universal properties within the Rényi framework. More significantly, this method can also be applied to modeling distributions in non-extensive statistical mechanics, such as the Tsallis ensemble~\cite{Tsallis1988}. Similar to the Rényi ensemble, the Tsallis ensemble depends on the average energy and has been shown to accurately describe the behavior of a wide range of strongly correlated, long-range Hamiltonians at finite temperatures, in particular in regimes where Boltzmann-Gibbs statistics fails due to the breakdown of ergodicity~\cite{Tsallis1988,Tsallis2016}.

Overall, our results showcase the potential and limitations of the Rényi ensemble as a framework for approximating thermal states and extracting universal critical properties of many-body systems. Our claim of the independence of the critical exponents of the 2D Ising transition on the Rényi index suggests that the $\alpha$-Rényi ensemble could serve as a rich playground for variational studies of finite-temperature phase transitions across various classical and quantum systems. Combining this approach with highly expressive models, such as RNNs~\cite{Raghu2017}, and techniques from recent tensor network and neural network simulations~\cite{Giudice2021,Giudice2024} may enable efficient and accurate variational studies for larger systems. With these advances, future research could extend this framework to the quantum computing realm, opening new avenues for the study of critical phenomena~\cite{TN_QuantumComputing_Miles_2024,TN_QuantumComputing_notMiles_2024}.

\section*{Open-Source Code}
Our Monte Carlo code, including code for the fixed point search, is made publicly available at 
"\url{https://github.com/andrewjreissaty91/ising_renyi_ensemble_MonteCarlo}", while details of the RNN implementation of Sec.~\ref{sec:RNN} can be found at 
"\url{https://github.com/andrewjreissaty91/ising\_renyi\_ensemble\_RNN}".

\section*{Acknowledgements}
We thank R. Wiersema, M. Duschenes, M. S. Moss, A. Orfi, M. Hibat-Allah, and A. Ijaz for their wisdom, insight, and valuable discussion. We also thank R. Brekelmans, R. G. Melko, L. Hayward, M. Reh, S. Czischek, and F. Oyedemi for their expert guidance and support throughout this project. We acknowledge the support of the Natural Sciences and Engineering Research Council of Canada (NSERC). JC acknowledges support from the Shared Hierarchical Academic Research Computing Network (SHARCNET), Compute Canada, and the Canadian Institute for Advanced Research (CIFAR) AI chair program. Resources used in preparing this research were provided, in part, by the Province of Ontario, the Government of Canada through CIFAR, and companies sponsoring the Vector Institute \url{www.vectorinstitute.ai/#partners}.

\appendix
\section{Mean-Field Details}
\label{appendix:mean_field_details}

\begin{figure}[htp]
\includegraphics[scale=0.53]{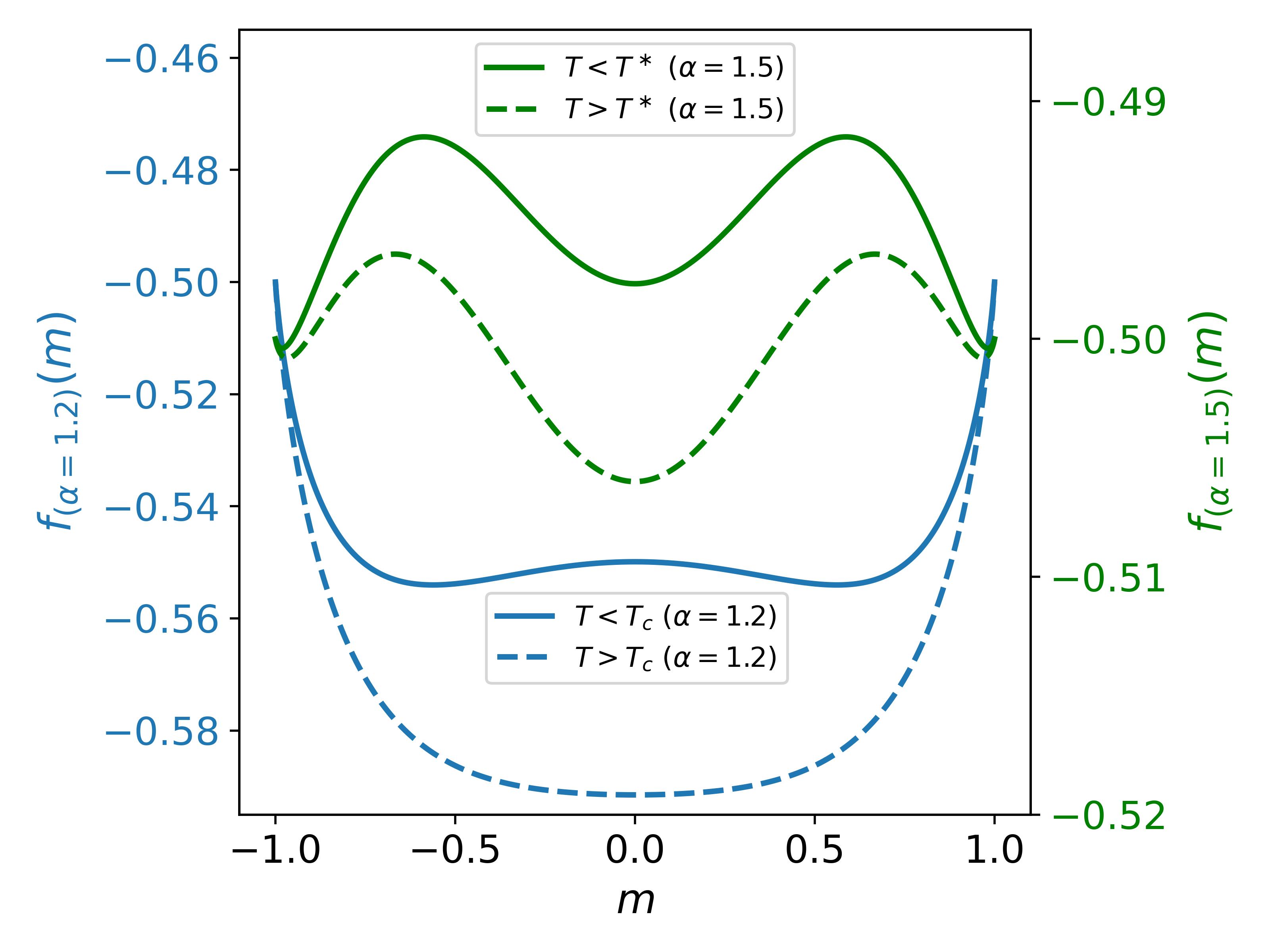}
\caption{\label{fig:epsart} Mean-field $\alpha$-Rényi free energy per spin (Eq.~(\ref{eq:free_energy_Renyi_Arovas_final_paper})) as a function of magnetization $m$ for a value of $\alpha$ for which the mean-field phase transition is continuous ($\alpha=1.2$) and  a value of $\alpha$ for which the transition is first-order ($\alpha=1.5$). The dashed and solid curves respectively showcase the free energy just above and below the given transition. At each temperature, the phase is determined by the value of $m$ that minimizes $f_\alpha$ globally, $\overline{m}$. Below the transition, the symmetry is broken ($\overline{m}\neq 0$) while above it, it is maintained ($\overline{m}=0$).}
\label{fig:mean_field_four_into_one_plot}
\end{figure}

As detailed in the main text, the mean-field product-state technique encapsulated by Eq.~(\ref{eq:rho_Arovas}) produces an expression for the Rényi free energy per spin $f_\alpha (m)$ detailed in Eq.~(\ref{eq:free_energy_Renyi_Arovas_final_paper}).  The expression $f_\alpha$ as a function of $m$ showcases the existence of a "threshold" value of the Rényi index $\alpha_{[c\rightarrow 1^\text{st}]} = \frac{\sqrt{13}-1}{2} \sim 1.303$ below which the Ising model in the $\alpha$-Rényi ensemble exhibits a continuous phase transition ($\alpha \in \left[1,\alpha_{[c\rightarrow 1^\text{st}]}\right]$), and above which it produces a first-order transition ($\alpha > \alpha_{[c\rightarrow 1^\text{st}]}$). This can be seen in Fig.~\ref{fig:mean_field_four_into_one_plot}, where $f_\alpha$ is plotted in the continuous regime above and below the critical temperature $T_c$, as well as in the first-order regime above and below the transition temperature $T^\ast$.

Let us now derive the threshold value $\alpha_{[c\rightarrow 1^\text{st}]}$. To that end, we need an analytic expression for $f_\alpha$ as a function of the order parameter $m$, in classic Landau tradition. We return to Eq.~(\ref{eq:free_energy_Renyi_Arovas_final_paper}), Taylor expand the logarithm about $m=0$ to $6^\text{th}$ order, and obtain
\begin{align} \label{eq:free_energy_6th_order}
\begin{split}
    f^{(6)}_\alpha & \approx -T\log 2 + \frac{1}{2}(\alpha T-qJ)m^2 \\
    & + \frac{\alpha T}{24}\left[(\alpha-2)(\alpha-3)-3\alpha(\alpha-1)\right] m^4 \\
    & + \frac{\alpha T}{720}\left[30\alpha^2 (\alpha-1)^2 - 15 \alpha(\alpha-1)(\alpha-2)(\alpha-3) \right.\\
    & + \left.(\alpha-2)(\alpha-3)(\alpha-4)(\alpha-5)\right] m^6,
\end{split}
\end{align}
where $f^{(6)}_\alpha$ has been defined as the $\mathcal{O}(m^6)$ approximation to $f_\alpha$. We can show that only even orders survive the Taylor expansion, fulfilling the Landau theory vision of having an analytic free energy that captures the symmetries of the Hamiltonian, in this case the $\mathbb{Z}_2$ spin-flip symmetry of the Ising model. We  justify ignoring higher orders than $\mathcal{O}(m^6)$ because in the vicinity of $\alpha_{[c\rightarrow 1^\text{st}]}$, the transition is either continuous or "nearly continuous", and so $m\sim0$ for all $T$ near $T_c$ or $T^\ast$. We also choose to restrict ourselves to $6^\text{th}$ order specifically because free energy expressions that capture first-order transitions tend to have five extrema (three minima and two maxima) near the transition temperature, and to produce this number of extrema, at minimum an $\mathcal{O}(m^6)$ expression is needed. By plotting $f_\alpha$ as a function of $m$ for values of $\alpha$ just above the numerically deduced $\alpha_{[c\rightarrow 1^\text{st}]}\sim 1.3$, we confirm that five extrema emerge in the interval $m\in[-1,1]$ when $T\sim T^\ast$, as can be seen in Fig.~\ref{fig:mean_field_four_into_one_plot}.

We find that the $f_\alpha^{(6)}$ approximation captures the exact value of $\alpha$ at which five extrema begin to emerge for $f_\alpha$ near the transition. When $f_\alpha$ has five extrema, $f_\alpha^{(6)}$ also has five extrema, although in the case of the latter, the extrema may extend beyond the interval $m\in[-1,1]$ depending on $\alpha$. In short, the $\mathcal{O}(m^6)$ analysis is sufficient for our goal. We now turn to Eq.~(\ref{eq:free_energy_6th_order}) to extract $\alpha_{[c\rightarrow 1^\text{st}]}$. Setting $\frac{\partial f^{(6)}_\alpha}{\partial m}=0$ and pulling out a factor $m$ which produces an extremum at $m=0$, the remaining extrema are the solutions of
\begin{equation} \label{eq:quadratic_equation}
    Am^4 + Bm^2 + C = 0,
\end{equation}
where we have defined
\begin{align}
    A & \equiv \frac{\alpha T}{120}\left[30\alpha^2 (\alpha-1)^2 - 15 \alpha(\alpha-1)(\alpha-2)(\alpha-3) \right.\nonumber\\
    & + \left.(\alpha-2)(\alpha-3)(\alpha-4)(\alpha-5)\right], \nonumber\\
    B & \equiv \frac{\alpha T}{6}\left[(\alpha-2)(\alpha-3)-3\alpha(\alpha-1)\right], \nonumber\\
    C & \equiv \alpha T-qJ. \nonumber
\end{align}
The solution to Eq.~(\ref{eq:quadratic_equation}) is
\begin{equation} \label{eq:quadratic_formula}
    m^2 = \frac{-B \pm \sqrt{B^2 - 4AC}}{2A}.
\end{equation}
In order for Eq.~(\ref{eq:quadratic_formula}) to be able to generate four real extrema (on top of the $m=0$ extremum discussed above), we must have $-B > 0$, while $m^2$ and the argument of the square root must also be positive. We find that for values of $\alpha$ just above the approximate value of $\alpha_{[c\rightarrow 1^\text{st}]} \sim 1.3$, the sign of $B^2-4AC$ depends on the temperature $T$. Thus, the emergence of five extrema is determined first by the sign of $-B$, and then by the choice of $T$. As such, setting $-B>0$ ends up being a sufficient condition for the derivation of the exact $\alpha_{[c\rightarrow 1^\text{st}]}$. We have
\begin{align}
    -\left[(\alpha-2)(\alpha-3)-3\alpha(\alpha-1)\right] & > 0, \nonumber\\
    \alpha^2 + \alpha - 3 & > 0, \nonumber\\
    \left(\alpha + \frac{1}{2}\right)^2 > \frac{13}{4} \longrightarrow \alpha & > \frac{\sqrt{13}-1}{2}.\nonumber
\end{align}
In other words, in order to have the possibility of five real extrema and thus a first-order transition, we must have $\alpha > \frac{\sqrt{13}-1}{2}$, which recovers equation Eq.~(\ref{eq:alpha_cont_to_first_order}) for $\alpha_{[c\rightarrow 1^\text{st}]}$.

Computing the exact $f_\alpha$ for any $\alpha$ above and below this exact threshold $\alpha$ confirms that $\alpha_{[c\rightarrow 1^\text{st}]} = \frac{\sqrt{13}-1}{2}$ separates the continuous and first-order regimes exactly. In addition, a related analysis using the $\mathcal{O}(m^4)$ Taylor expansion of $f_\alpha$, one that we do not detail here, can be performed, and we find that it also produces the same exact result for $\alpha_{[c\rightarrow 1^\text{st}]}$.

\section{The $\alpha$-Rényi Ensemble Fixed Point} 
\label{appendix:attractive_fixed_point}
Let us now consider the Rényi ensemble at $\alpha=2$. We wish to study the nature of its fixed point. We define the function $f_2\left(\bar{E}\right)$ as
\begin{equation} \label{eq:f_2_of_Ebar}
    f_2\left(\bar{E}\right) =  \frac{\sum\limits_{j=0}^{n_\beta-1} N_j E_j \left[1 - \frac{1}{2}\beta(E_j - \Bar{E})\right]}{\sum\limits_{j=0}^{n_\beta-1} N_j \left[1 - \frac{1}{2}\beta(E_j - \Bar{E})\right]},
\end{equation}
which corresponds to the left-hand side of Eq.~(\ref{eq:fixed_point_equation})
with $\alpha=2$. At the fixed point $\Bar{E} = \Bar{E}_\text{fp}$, we have $f_2\left(\Bar{E}_\text{fp}\right)=\Bar{E}_\text{fp}$. Now, in order to prove that the fixed point is attractive, we would have to show that
\begin{equation} \label{eq:condition3_attractive_fixed_point}
    \left|f_2\left(\Bar{E}_\text{fp}+d\Bar{E}\right) - f_2\left(\Bar{E}_\text{fp}\right)\right| < \left|d\Bar{E}\right|,
\end{equation}
where we have defined $d\Bar{E}$ as an infinitesimal perturbation away from the fixed point. Given Eq.~(\ref{eq:f_2_of_Ebar}), we can write
\begin{align}
    f_2&\left(\Bar{E}_\text{fp}+d\Bar{E}\right) = \nonumber\\
    &\frac{\sum\limits_{j=0}^{n_\beta-1} N_j E_j \left[1 - \frac{1}{2}\beta(E_j - \Bar{E}_\text{fp})\right] + \frac{1}{2}\beta d\bar{E} \sum\limits_{j=0}^{n_\beta-1} N_j E_j}{\sum\limits_{j=0}^{n_\beta-1} N_j \left[1 - \frac{1}{2}\beta(E_j - \Bar{E}_\text{fp})\right] + \frac{1}{2}\beta d\bar{E} \sum\limits_{j=0}^{n_\beta-1} N_j} \\ \label{eq:XplusX_YplusY}
    & \equiv \frac{X + X'}{Y + Y'},
\end{align}
where the variables $X\equiv\sum\limits_{j=0}^{n_\beta-1} N_j E_j \left[1 - \frac{1}{2}\beta(E_j - \Bar{E}_\text{fp})\right]$, $X'\equiv\frac{1}{2}\beta d\bar{E} \sum\limits_{j=0}^{n_\beta-1} N_j E_j$, $Y\equiv\sum\limits_{j=0}^{n_\beta-1} N_j \left[1 - \frac{1}{2}\beta(E_j - \Bar{E}_\text{fp})\right]$ and $Y'\equiv\frac{1}{2}\beta d\bar{E} \sum\limits_{j=0}^{n_\beta-1} N_j$ have been defined to supplement the subsequent analysis. In all the above equations, we have assumed that a perturbation of the fixed point $\Bar{E}_\text{fp}$ by an infinitesimal quantity $d\Bar{E}$ does not change the set of $n_\beta-1+1=n_\beta$ energies that satisfy the constraint in Eq.~(\ref{eq:alphaRenyiEnsConstr}) when $\Bar{E} = \Bar{E}_\text{fp}$.

Next, we consider Eq.~(\ref{eq:condition3_attractive_fixed_point}). In the language of $X$ and $Y$, if the fixed point is attractive, it must mean that
\begin{equation} \label{eq:condition4_attractive_fixed_point}
    \left| \frac{X+X'}{Y+Y'} - \frac{X}{Y} \right| < \left| d\bar{E} \right|.
\end{equation}
Some algebraic work yields
\begin{align}
    &\left| \frac{X+X'}{Y+Y'} - \frac{X}{Y} \right| = \left| \frac{\frac{1}{2}\beta\sum\limits_{j=0}^{n_\beta-1} N_j(YE_j - X)d\Bar{E}}{Y^2 + Y\frac{1}{2}\beta\sum\limits_{j=0}^{n_\beta-1} N_j d\Bar{E}} \right| \nonumber\\ \label{eq:condition5_attractive_fixed_point}
    &\approx \underbrace{\left| \frac{\frac{1}{2}\beta\sum\limits_{j=0}^{n_\beta-1} N_j(YE_j - X)}{Y^2} \right|}_{\equiv A} \left| d\Bar{E} \right|,
\end{align}
where the fraction $A$, which we can write as
\begin{align}
    &A = \nonumber\\
    &\left|\frac{\frac{1}{2}\beta \sum\limits_{j=0}^{n_\beta-1} \sum\limits_{k=0}^{n_\beta-1} N_j N_k \left[ 1-\frac{1}{2}\beta(E_j-\bar{E}_\text{fp})\right](E_k-E_j)}{\sum\limits_{j=0}^{n_\beta-1} \sum\limits_{k=0}^{n_\beta-1} N_j N_k \left[1-\frac{1}{2}\beta(E_j-\bar{E}_\text{fp})\right]\left[ 1-\frac{1}{2}\beta(E_k-\bar{E}_\text{fp})\right]}\right| \nonumber\\ 
    &=  \left| \frac{\frac{1}{2}\beta\sum\limits_{k=0}^{n_\beta-1} N_k (E_k - \bar{E}_\text{fp})}{\sum\limits_{k=0}^{n_\beta-1} N_k \left[1-\frac{1}{2}\beta(E_k - \bar{E}_\text{fp})\right]} \right| \nonumber\\
    \label{eq:fractionA_lessThanOrGreaterThan1}
    &= \left| \frac{\frac{1}{2}\beta\sum\limits_{k=0}^{n_\beta-1} N_k (E_k - \bar{E}_\text{fp})}{\sum\limits_{k=0}^{n_\beta-1} N_k - \frac{1}{2}\beta\sum\limits_{k=0}^{n_\beta-1} N_k (E_k - \bar{E}_\text{fp})}\right|,
\end{align}
must be less than $1$ in order for the fixed point to be attractive. In the high-temperature limit, $\beta$ tends to zero and we have $\sum\limits_{k=0}^{n_\beta-1} N_k \gg \mathcal{O}(\beta)$, producing $A<1$ and thus an attractive fixed point. In the opposite limit $T\rightarrow 0$, the ground state fixed point $\Bar{E}_\text{fp}=E_0$ must be attractive as well since the sums in Eq.~(\ref{eq:fractionA_lessThanOrGreaterThan1}) have only one term $E_k - \Bar{E}_\text{fp}=E_0 - E_0 = 0$, so at low temperatures, the ground state fixed point, which our 2D Monte Carlo simulations of the Ising model indeed find, is attractive as well with $A<1$. We also showed in Sec.~\ref{sec:1D_exact_solution_final_paper} that in the 1D Ising model in the generalized $\alpha$-Rényi ensemble, there are an infinite number of lower energy fixed points $E_\text{fp}=E_0,E_1,E_2,...$ that can be found in the thermodynamic limit for any $T\in \left[0,\frac{\alpha-1}{\alpha} 4J\right)\bigg|_{\alpha=2} = \left[0,2J\right)$ such that $E_j \le E_\text{fp}$ for all allowed energies $\{E_j\}$. For each of those those fixed points, we have $\sum\limits_{k=0}^{n_\beta-1} N_k (E_k - \bar{E}_\text{fp}) \le 0$, producing $A<1$ in Eq.~(\ref{eq:fractionA_lessThanOrGreaterThan1}) and thus a large set of attractive fixed points.

In our exact and Monte Carlo simulations of the 1D and 2D Ising models with no external field, we observe that the maximum energy fixed point, which we explicitly target, is also attractive for all temperatures and values of the Rényi index $\alpha > 1$ that we test.

\section{Shifted $|m|$ vs $T$ curves} 
\label{appendix:m_vs_T_overlap}

\begin{figure}[htp]
\includegraphics[scale=0.53]{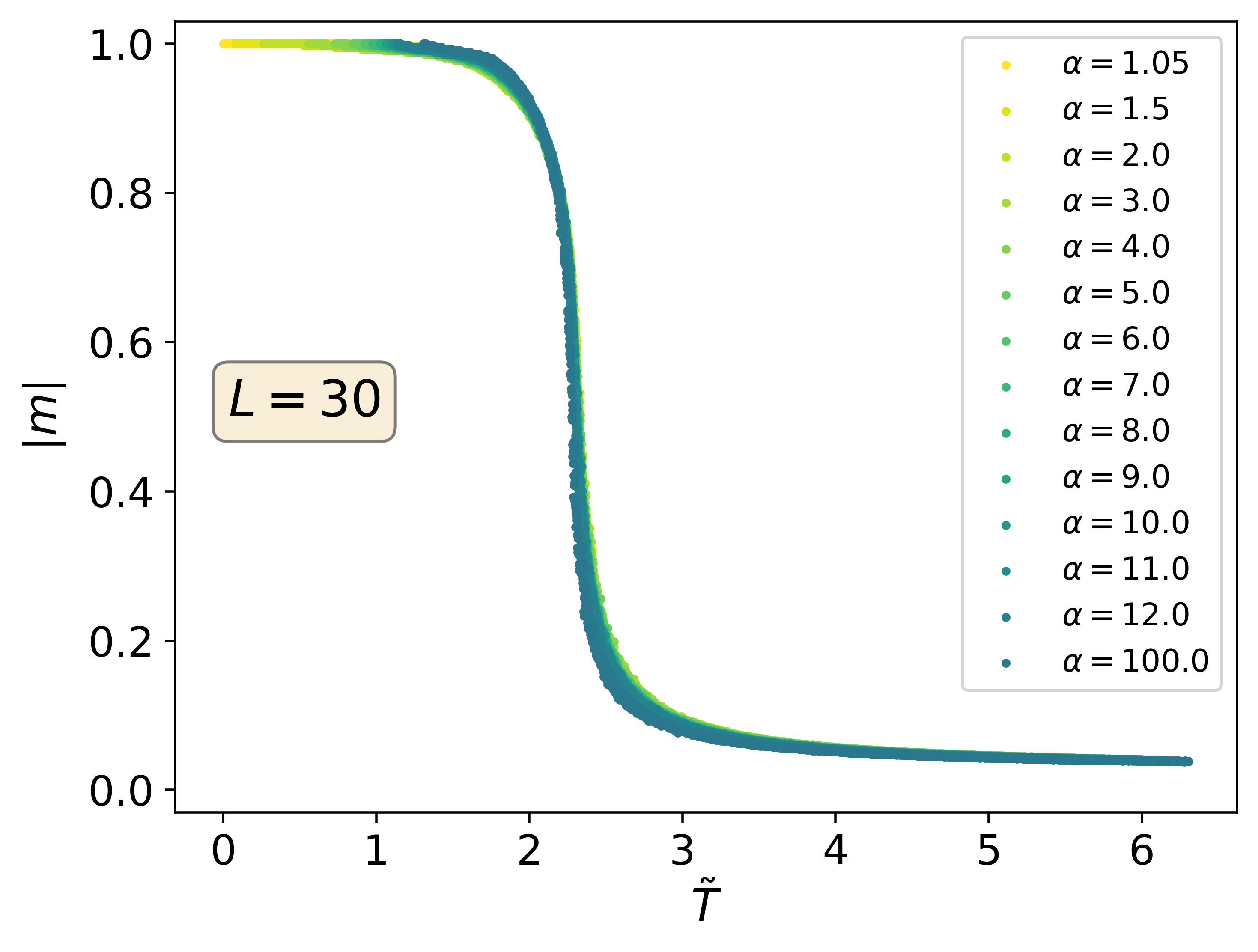}
\caption{\label{fig:epsart} Results of Fig.~\ref{fig:m_vs_T_MonteCarlo_L30_allAlpha} with $|m|$ plotted this time as a function of $\Tilde{T}$, defined as $T\rightarrow \Tilde{T}=T+T_c(\alpha=1.05)-T_c(\alpha)$. Here, we denote $T_c$ as $T_c(\alpha)$ to emphasize that it is a function of $\alpha$, so the shift amount $T_c(\alpha=1.05)-T_c(\alpha)$ is different for each curve in Fig.~\ref{fig:m_vs_T_MonteCarlo_L30_allAlpha}.}
\label{fig:m_vs_T_shiftedCurves}
\end{figure}

\begin{figure*}
\includegraphics[scale=0.85]{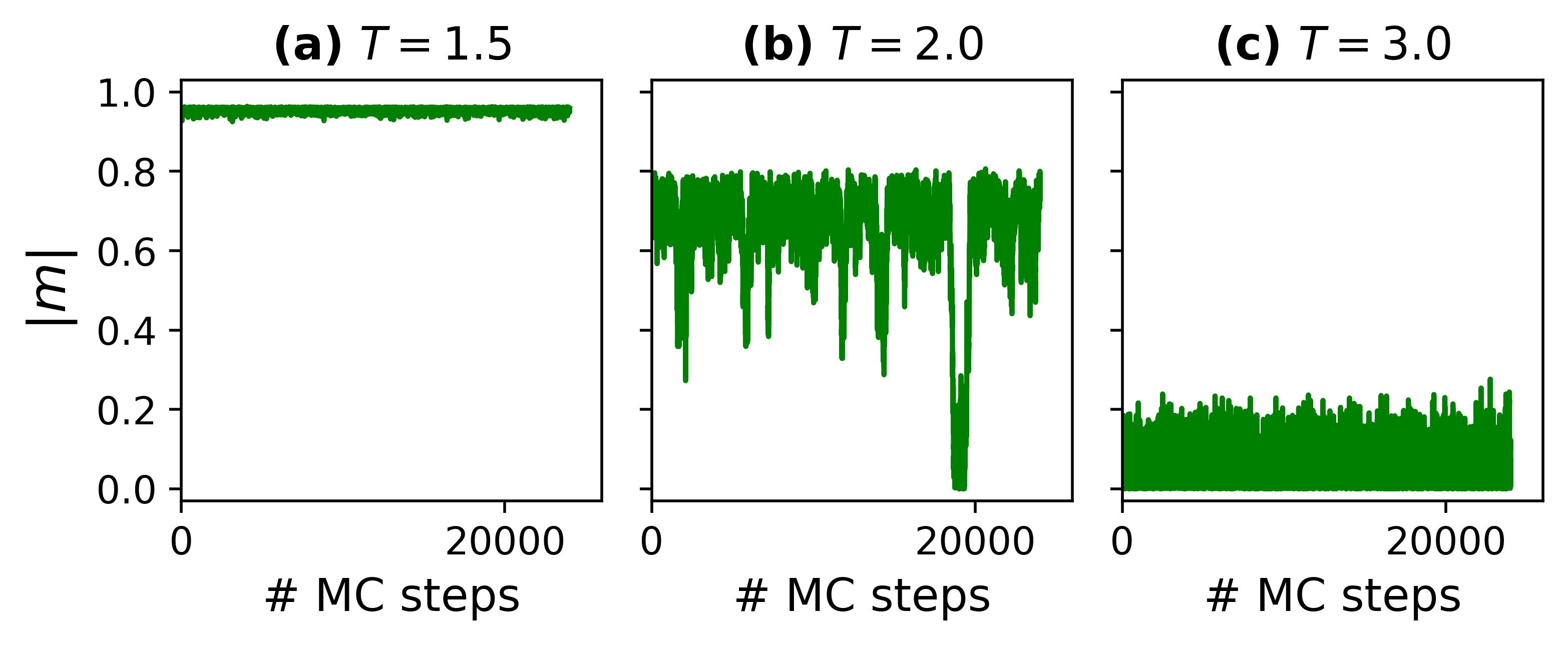}
\vspace{0pt}
\caption{\label{fig:epsart} Absolute value of the magnetization $|m|$ as a function of Monte Carlo time (one MC step = one full sweep of the lattice) for a $40\times40$ 2D Ising model in the $2$-Rényi ensemble at \textbf{(a)} T = 1.5 (below $T_c$), \textbf{(b)} T = 2.0 (approximately $T_c$), \textbf{(c)} $T=3.0$ (above $T_c$). All three simulations are conducted after the fixed point is found, and equilibrium is reached relatively quickly, with each simulation essentially being a continuation of the final run of the fixed point search (see Sec.\ref{sec:Monte_Carlo} for more details on the specific procedure). Averages are computed by discarding half the Monte Carlo steps  in these plots (considered "thermalization time", even though the target distribution is reached relatively quickly). Using the binning technique, the autocorrelation time for the $T\sim T_c$ simulation in \textbf{(b)} is found to be $\tau \sim 80$ Monte Carlo steps, with $\tau$ decreasing as $T$ moves away from $T_c$. We note that $24000$ MC steps are taken overall in all three simulations.}
\label{fig:m_vs_steps}
\end{figure*}

The results of Fig.~\ref{fig:m_vs_T_MonteCarlo_L30_allAlpha} suggest that the shape of the $|m|$ vs $T$ curves at fixed linear system size $L$ through the transition temperature $T_c$ may be $\alpha$-independent, so long as $L$ is large enough for the curves to appear continuous. In other words, the derivative $\frac{\partial |m|}{\partial T}$ evaluated in the vicinity of $T\sim T_c(\alpha)$ may be unchanged across all $\alpha\ge1$, as the curves look qualitatively similar through the critical point (we write $T_c\rightarrow T_c(\alpha)$ to emphasize that $T_c$ is a function of $\alpha$). Estimating derivatives based on Monte Carlo results is difficult. Instead, we perform a simple rudimentary test of our hypothesis by shifting the $|m|$ vs $T$ curves in Fig.~\ref{fig:m_vs_T_MonteCarlo_L30_allAlpha} to the right to overlap with the $\alpha=1.05$ curve. We accomplish this by mapping $T$ to a shifted temperature $\Tilde{T}$ as per $T\rightarrow \Tilde{T}=T+T_c(\alpha=1.05)-T_c(\alpha)$ and plotting $|m|$ as a function of $\Tilde{T}$. Since $T_c(\alpha)$ depends on $\alpha$, each curve is shifted by a different amount. We note that the set of $\{T_c(\alpha)\}$ used for the shift are the critical temperatures as extracted by the data collapse tuning process described in Sec.~\ref{sec:Monte_Carlo}.

The results are shown in Fig.~\ref{fig:m_vs_T_shiftedCurves}, and the curves overlap well through their respective critical temperatures, indicating that $\frac{\partial |m|}{\partial T}$ may well be independent of $\alpha$ near $T\sim T_c(\alpha)$. Since critical exponents depend on the behavior of observables such as $|m|$ and other free energy derivatives as $T$ approaches $T_c$ from above and below, this is yet further evidence supporting our claim that the critical exponents of the 2D Ising transition in the $\alpha$-Rényi ensemble are independent of $\alpha$, on top of the numerical evidence which comes from the data collapse of Fig.~\ref{fig:dataCollapse_FourPlots}. If the claim ultimately proves true, it would be in agreement with the $\alpha$-independence of the exponents in mean-field theory seen in  Sec.~\ref{sec:mean-field_final_paper}.

\section{Monte Carlo Statistics and Thermalization} 
\label{appendix:MC_stats}

In Fig.~\ref{fig:m_vs_steps}, we showcase the absolute value of the magnetization $|m|$ as a function of Monte Carlo steps, where one step corresponds to a full sweep of the lattice, for three different temperature simulations of a $40\times40$ 2D Ising model in the $2$-Rényi ensemble ($\alpha=2$). The simulations were performed below the critical point ($T=1.5$,  Fig.~\ref{fig:m_vs_steps}(a)), near criticality ($T = 2.0 \sim T_c$, Fig.~\ref{fig:m_vs_steps}(b)), and above $T_c$ ($T=3.0$, Fig.~\ref{fig:m_vs_steps}(c)). All three simulations are conducted after finding the fixed point using the method described in Sec.~\ref{sec:Monte_Carlo}. The equilibrium is generated relatively quickly, aided by each simulation essentially being a continuation of the final run of the fixed point search. Clearly, the fluctuations are significantly larger at criticality than away from $T_c$ due to the effects of longer autocorrelation times and critical slowing down \cite{BarkemaNewmanBook}, but an equilibrium is realized nonetheless in Fig.~\ref{fig:m_vs_steps}(b). The autocorrelation time $\tau$ for the critical point simulation (denoted $\tau_c$) is around $\tau_c \sim 80$ sweeps of the lattice, calculated as per Eq.~(23) of Ref.~\cite{Troyer_autocorrTime_2010}. For values of the Rényi index approximately satisfying $\alpha \in [2,5]$, $\tau_c$ is approximately constant, but it then experiences a steady increase as $\alpha$ is increased beyond $5$, e.g. $\tau_c \sim 150$ sweeps when $\alpha = 8$. Below $\alpha=2$, $\tau_c$ decreases, with $\tau_c \sim 40$ sweeps for $\alpha=1.5$. As expected, we find that $\tau$ decreases as $T$ moves away from $T\sim T_c$, for all $\alpha$. We note that the number of bins and Monte Carlo steps per bin used to calculate averages for all the Monte Carlo simulations we conduct in this paper were dependent on system size, and we chose them in a trial-and-error fashion in such a way to make the binning technique converge.

\section{Susceptibility} 
\label{appendix:Chi_Binder}

For classical spin models in the Gibbs state, the magnetic susceptibility per spin can be computed via 
\begin{equation} \label{eq:chi}
    \chi \equiv N \cdot \frac{\langle m^2\rangle - \langle m \rangle^2}{T}
\end{equation}
where $m$ is the magnetization per spin \cite{BarkemaNewmanBook}. It is a second derivative of the free energy, and thus it can be compactly expressed in terms of the Gibbs state partition function \cite{GoldenfeldBook,BarkemaNewmanBook}. For the generalized $\alpha$-Rényi ensemble, Eq.~(\ref{eq:chi}) does not necessarily hold. However, because of the success of the data collapse approach of Sec.~\ref{sec:Monte_Carlo} applied to our Rényi ensemble data, we are motivated to trial the expression for $\chi$ above on the statistics generated by the Rényi ensemble. Specifically, since $\chi$ is known to diverge at a Gibbs state critical point in the thermodynamic limit, we now look for hallmarks of this divergence for $\alpha > 1$.

\begin{figure}[htp]
\includegraphics[scale=0.57]{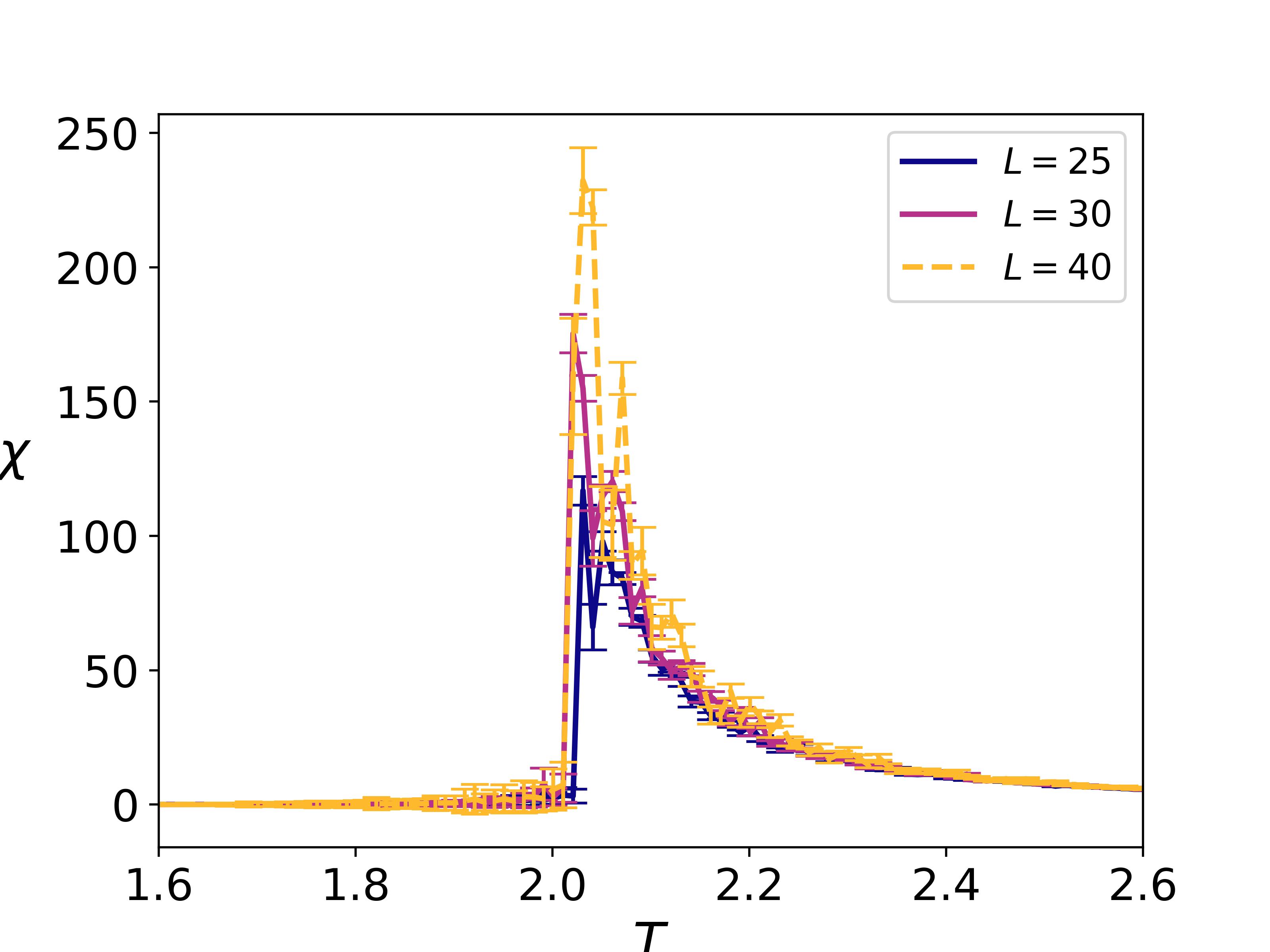}
\caption{\label{fig:epsart} Magnetic susceptibility per spin $\chi$ as a function of $T$ for the 2D Ising model in the $2$-Rényi ensemble at three different system sizes. $\chi$ is computed using the Gibbs state form Eq.~(\ref{eq:chi}). While this form does not necessarily apply to the generalized Rényi ensemble beyond the limit $\alpha\rightarrow1$, it still produces data that peaks in the vicinity of the value of $T_c$ that is extracted using the data collapse approach of Fig.~\ref{fig:dataCollapse_FourPlots}. The error bars are generated using the Monte Carlo error approach of Ref.~\cite{BeccaSorellaBook}. We must stress that these are minimum errors that do not take the errors associated with the fixed point search into account (see Sec.~\ref{sec:Monte_Carlo} for more on this specific point).}
\label{fig:chi_vs_T_alpha2.0}
\end{figure}

In Fig.~\ref{fig:chi_vs_T_alpha2.0}, we plot $\chi$ as a function of $T$ at three different system sizes for the 2D Ising model in the $2$-Rényi ensemble. $\chi$ seems to peak in the vicinity of $T\sim 2.0$ which is approximately the value of $T_c$ extracted using our data collapse approach for $\alpha=2$ in Fig.~\ref{fig:Tc_vs_alpha}. The many sources of error, especially the error associated with finding the Rényi ensemble fixed point, are such that the true error bars in the vicinity of the transition are relatively large. We must stress that the bars we depict in Fig.~\ref{fig:chi_vs_T_alpha2.0} are minimum errors, computed using traditional Monte Carlo methods \cite{BeccaSorellaBook}, and that do not take the fixed point search errors into account. Nevertheless, it seems likely based on Fig.~\ref{fig:chi_vs_T_alpha2.0} that $\chi$ will diverge in the limit $N\rightarrow\infty$, and it seems as if it will do so at a temperature near $T\sim 2.0$, lending further credence to our data collapse approach which fixes the critical exponents $\beta$ and $\nu$ to their Gibbs state values. Though we do not show the results at other values of $\alpha$, similar peaks in $\chi$ are produced as $\alpha$ is varied, all in the vicinity of the critical temperatures extracted by our data collapse.

\clearpage
\bibliography{MainPaper}{}

\end{document}